\documentclass{elsarticle}
\usepackage[utf8]{inputenc}
\usepackage[T1]{fontenc}
\usepackage{amsmath}
\usepackage{amssymb}

\usepackage[table]{xcolor}

\usepackage{todonotes}
\presetkeys{todonotes}{inline}{}

\usepackage{hyperref}

\usepackage[lighttt]{lmodern}
\usepackage{listings}
\let\code\texttt

\lstnewenvironment{lstfloat}[1][]%
{\lstset{basicstyle=\scriptsize\ttfamily,basewidth=0.5em,frame=TBlr,float=hbtp,#1}}{}
\lstnewenvironment{lstdisplay}[1][]%
{\lstset{basicstyle=\small\ttfamily,xleftmargin=1em,basewidth=0.5em,frame=none,#1}}{}

\let\vec\mathbf
\providecommand{\abs}[1]{\left|#1\right|}
\providecommand{\norm}[1]{\lVert#1\rVert}

\newcounter{bla}

\journal{Computer Physics Communications}

\begin{document}

\begin{frontmatter}



\title{Fast, feature-rich weakly-compressible SPH on GPU: coding strategies and compiler choices}


\author[a]{Giuseppe Bilotta\corref{author}}
\author[b]{Vito Zago}
\author[a,c]{Alexis H\'erault}
\author{Hendrik D. van Ettinger}
\author[b]{Robert A. Dalrymple}

\cortext[author] {Corresponding author.\\\textit{E-mail address:} giuseppe.bilotta@ingv.it}
\address[a]{Osservatorio Etneo, Istituto Nazionale di Geofisica e Vulcanologia, Catania, Italy}
\address[b]{Dept. of Civil and Environmental Engineering, Northwestern University, 2145 Sheridan Road, Evanston, IL 60208, USA}
\address[c]{Laboratoire Mod\'elisation math\'ematique et numerique,
	Conservatoire National des Arts et M\'etiers, 292 Rue Saint-Martin, Paris 75003, France}

\begin{abstract}
GPUSPH was the first implementation of the weakly-compressible Smoothed Particle Hydrodynamics
method to run entirely on GPU using CUDA. Version 5, released in June 2018, features a radical
restructuring of the code, offering a more structured implementation of several features
and specialized optimization of most heavy-duty computational kernels.
While these improvements have led to a measurable performance boost
(ranging from 15\% to 30\% depending on the test case and hardware configuration),
it has also uncovered some of the limitations of the official CUDA compiler (\code{nvcc})
offered by NVIDIA, especially in regard to developer friendliness.
This has led to an effort to support alternative compilers, particularly Clang,
with surprising performance gains.
\end{abstract}


\end{frontmatter}


\section{Introduction}

Smoothed Particle Hydrodynamics (SPH) is a Lagrangian, meshless numerical method for computational fluid dynamics
originally created for astrophysics~\cite{Lucy:1977,gingold-monaghan-1977},
and that has since grown to cover a wide range of fields~\cite{monaghan_2005} thanks to its ability to handle complex flows \cite{zago_aog_2017}.
The Lagrangian, meshless nature of the method makes it particularly apt for free surface flows,
violent flows, temperature-dependent fluids and non-Newtonian fluids~\cite{Cleary:1999,Monaghan:2005,bilotta_2016}.

One feature that makes the standard weakly-compressible form of SPH (WCSPH)
particularly attractive from a computational point of view is the embarrassingly parallel
nature of the method: the time evolution of each particle can be computed directly
from the properties of the particle itself and those of its immediate neighborhood,
without requiring the solution of any linear system,
leading to straightforward implementation on massively parallel hardware.

In the last decades, graphic processing units (GPUs) have become a cheap alternatives
to traditional CPU clusters as consumer-friendly parallel computing hardware~\cite{gems2,harada2007}.
The mass adoption of GPUs as computing solutions has been spearheaded by
NVIDIA with CUDA, a runtime library with an associated single-source extension
to the C++ programming language that makes it relatively easy to write software
that can run on their GPUs~\cite{cuda1}.

GPUSPH was the first implementation to leverage the capabilities of CUDA with an implementation
of WCSPH that could run entirely on NVIDIA GPUs \cite{herault_sph_2010}.
Throughout its history, performance has always been a priority in the development of GPUSPH,
hence the choice to focus on a GPU-only implementation, and its expansion to multi-GPU \cite{rustico_advances_2014}
and multi-node (GPU clusters) \cite{rustico_multi-gpu_2014} systems.

The growing support for more recent revisions of the C++ standard in CUDA
has allowed us to improve the design of the GPUSPH codebase and its performance
without any sacrifice to functionality, and in fact making it easier to implement
new features. The main downside in these advances has been the limitations
imposed by the NVIDIA CUDA compiler, that has led us to explore
alternative compilers, with surprising results.

The paper is written to be as self-contained as possible,
but it does assume that the reader is already familiar with the fundamentals of C++ programming~\cite{stroustroup2014}.

In the first part of this paper we will discuss briefly the features offered by GPUSPH
and how the development goals of the project affect their implementation
(section~\ref{sec:features}),
the technical choices made to reduce the maintenance cost of the code base
and the associated performance benefits and downsides
(sections~\ref{sec:cpp11} to~\ref{sec:down}).
Many of the techniques discussed in this article were first introduced in GPUSPH version~5,
and were partially presented in our previous works~\cite{bilotta_spheric_2019,bilotta_design_2018}.
Here we cover these topics in more detail,
and include several refinements never discussed before,
mainly aimed at improving code readability leveraging the expressivity of more modern revisions of the C++ standard.
In the second part of the paper (from section~\ref{sec:clang})
we discuss the astonishing side effects of introducing support for alternative toolchains,
and how it helped identify additional bottlenecks in our code,
the resolution of which led to significant performance gains to the benefits of industrial applications of GPUSPH
(section~\ref{sec:benchmarks}).
All of the improvements discussed in this paper are available on the public version of GPUSPH,
which can be obtained from the project's GitHub repository\footnote{\url{https://github.com/GPUSPH/gpusph}},
and will be officially released as part of version~6 of the software.

\section{Weakly-compressible SPH}

\subsection{A lightning introduction to the method}

As a Lagrangian method for computational fluid dynamics, weakly-compressible SPH
is designed to solve the equations for the continuity of mass
\begin{equation}\label{eq:mass-cont}
\frac{D\rho}{Dt} =  -\rho \nabla\cdot\vec u
\end{equation}
and momentum (Navier--Stokes equations)
\begin{equation}\label{eq:ns}
\frac{D \vec u}{Dt} = -\frac{\nabla P}{\rho} + \frac{1}{\rho} \nabla\cdot(\mu \nabla \vec u) + \vec g
\end{equation}
where $\rho$ represents the density, $\vec u$ the velocity, $P$ the pressure,
$\mu$ the dynamic viscosity, $\vec g$ the external body forces (e.g. gravity),
and $D/Dt$ is the Lagrangian (total) derivative with respect to time.

The system of equations is closed by an equation of states that relates
the pressure $P$ to the density $\rho$, typically Cole's \cite{cole, batchelor}
equation of state
\[
P(\rho) = B \left( \left(\frac{\rho}{\rho_0}\right)^\gamma - 1\right)
\]
where $\rho_0$ is the at-rest density for the fluid, $\gamma$ the polytropic constant,
and $B$ a coefficient related to the at-rest sound speed $c_0$
by $B = \rho_0 c_0^2 / \gamma$.
Weak compressibility is achieved under the assumption that $c_0 > 10 U$,
where $U$ is the maximum flow velocity, which ensures that relative density variations
will remain below $1\%$.

Although the physical speed of sound of the fluid would be sufficient to guarantee
the weak-compressibility condition in many applications with subsonic flows (Mach number < 0.1),
the spatial discretization of WCSPH (that will be presented momentarily) is often paired with an explicit integration scheme,
for which the physical speed of sound would result in prohibitively small time-steps.
In practical applications of WCSPH a fictitious sound speed is usually preferred,
chosen lower than the physical one, but high enough to maintain the weakly-compressible regime.

In this case, in the computation of $U$ (and thus $c_0$),
one should take into account not only the actual velocity
experienced by the particles due to the dynamics, but also the hydrostatic condition,
defined by the theoretical free-fall velocity experienced by a particle dropping
from the maximum fluid height to the lowest point:
assuming $g$ is the magnitude of $\vec g$ and $H$ is the maximum distance that
can be travelled by a particle in the direction of $\vec g$, the hydrostatic condition
can be computed as $\sqrt{2gH}$.

With SPH, the computational domain $\Omega$ is discretized by a set of particles
that act as interpolation nodes, but are free to move with respect to each other.
Any field $f$ is then discretized by representing it as
as a convolution with Dirac's $\delta$ distribution
$f(\vec x) = \int_\Omega f(\vec y) \delta(\vec y - \vec x) d \vec y$,
approximating Dirac's distribution by means of a family
of \emph{smoothing kernels} $W(\vec r, h)$ parametrized by the \emph{smoothing length}
$h$ in such a way that $\lim_{h\to0} W = \delta$ in the sense of distributions,
and finally discretizing the integral as a summation over all the particles:
\begin{equation}\label{eq:sph-field}
f(\vec x) \simeq \sum_j f(\vec x_j) W(\vec x_j - \vec x, h) V_j.
\end{equation}
where $V_i$ represents the volume of particle $i$, and is frequently expressed
in terms of its mass and density as $V_i = m_i/\rho_i$.

The smoothing kernel is usually chosen radial (i.e. depending only on $r = \norm{\vec r}$),
with compact support
(specifically, there exists $k > 0$ such that $W(\vec r, h) = 0 \forall \vec r \text{ s.t. } r > kh$)
and unitary (i.e. such that $\int_\Omega W(\vec r, h) d\vec r = 1$).

The compact support implies that the summation \eqref{eq:sph-field}
only extends to the \emph{neighborhood} of $\vec x$ of radius $kh$ (called the \emph{influence radius} of the kernel).
The radial symmetry implies that $W(\vec r, h) = \omega(r, h)$ for some function $\omega$,
and that the kernel gradient can be written as
$\nabla W(\vec r , h) = \vec r F(r, h)$ where $F(r, h) \triangleq (1/r) \partial \omega/\partial r$,
which is particularly convenient when $F$ can be written analytically without an explicit division by $r$,
improving numerical stability when $r$ may become vanishingly small.

Using the standard SPH notation $\vec x_{ij} = \vec x_i - \vec x_j$, $W_{ij} = W(\vec x_{ij}, h)$,
$F_{ij} = F(\abs{\vec x_{ij}}, h)$, and $f_{ij} = f(\vec x_i) - f(\vec x_j)$ for any other field $f$,
the SPH discretization of the gradient of a field $f$
at the position of particle $j$ which is far from the boundary of the domain can be written as
\begin{equation}\label{eq:sph-grad}
\nabla f(\vec x_i) \simeq \sum_j f(\vec x_j) F_{ij} V_j \vec x_{ij}
\end{equation}
although symmetrized expressions, obtained by adding/subtracting the gradient of a constant field,
are preferred, leading to expressions such as
\begin{align*}
\nabla f(\vec x_i) & \simeq \sum_j \left(f(\vec x_i) + f(\vec x_j) \right) F_{ij} V_j \vec x_{ij}\\
\text{or}\\
\nabla f(\vec x_i) & \simeq \sum_j f_{ij} F_{ij} V_j \vec x_{ij}.
\end{align*}

The choice of the form for the discretization of the gradient leads to a variety of different SPH
formulations \cite{monaghan_2005, grenier_2009, hu_adams, colagrossi_landrini_2003}.
For example, a common formulation, following Monaghan's ``golden rule'',
expresses equations \eqref{eq:mass-cont} and \eqref{eq:ns} in discrete form as
\begin{align}
\frac{D\rho_i}{Dt} &= -\sum_j m_j \vec u_{ij} \cdot \vec x_{ij} F_{ij}\label{eq:mass-cont-sph1}\\
\frac{D \vec u_i}{Dt} &= -\sum_j \left(
\left( \frac{P_i}{\rho_i^2} + \frac{P_j}{\rho_j^2}\right) \vec x_{ij}
+ 2K \frac{m_j}{\rho_i \rho_j} \bar\mu_{ij}
\frac{\vec x_{ij} \cdot \vec u_{ij}}{\vec x_{ij}^2} \vec x_{ij} \right) F_{ij} + \vec g\label{eq:ns-sph1}
\end{align} 
while an alternative formulation taking ideas from Landrini \cite{colagrossi_landrini_2003} and Morris \cite{morris_1997} gives:
\begin{align}
\frac{D\rho_i}{Dt} &= -\sum_j \frac{\rho_i}{\rho_j} m_j \vec u_{ij} \cdot \vec x_{ij} F_{ij}\label{eq:mass-cont-sph2}\\
\frac{D \vec u_i}{Dt} &= -\sum_j \left(
\frac{P_i + P_j}{\rho_i\rho_j} \vec x_{ij}
+ 2 \frac{\rho_i \rho_j}{\rho_i + \rho_j} \bar\mu_{ij} \vec u_{ij} \right) F_{ij} + \vec g\label{eq:ns-sph2}
\end{align} 

Additionally, dissipative terms may be added to both the
momentum \cite{monaghan_2005} and mass \cite{molteni_colagrossi_2009, marrone_2011}
conservation equations, to smooth out numerical noise.
The expression for the smoothing kernel and its gradient can also be replaced by
corrected versions that improve the consistency and\slash or conservative properties
of the discretized operators~\cite{chen_1999,vila_1999,guilcher_2007,zago_ccsph}.

\subsection{From theory to implementation}
\label{sec:sph-impl}

Since the time derivatives of the particle properties in WCSPH can be computed from
the properties of the particle and its neighbors, the method lends itself naturally to parallelization,
especially when coupled with an explicit integration scheme.

On stream processing hardware such as GPUs, there is a natural mapping between work-items and particles
that has been exploited for SPH implementation even before the birth of GPGPU-enabled hardware~\cite{harada2007}.
With this focus, we can talk about the \emph{central particle} (the one being processed by the work-item),
and its \emph{neighbors}, the particles contained in the influence sphere of the central particle and that
participate in the summations on the right-hand side of equations \eqref{eq:mass-cont-sph1}--\eqref{eq:ns-sph2}
(or any other discretized formulation of choice).

The main iteration of most numerical implementations of WCSPH thus follows a scheme like the following:
\begin{description}
\item[neighbors search] used to identify the neighbors of each particle;
\item[computation of time derivatives] computing $D\rho/Dt$, $D\vec u/Dt$, etc for each particle;
\item[integration] computing the new density, velocity and position.
\end{description}

The computation of time derivatives and their integration may happen multiple times per integration step,
depending on the adopted scheme (e.g. once for a simple forward Euler integration scheme,
twice in a predictor\slash corrector scheme, four times in a Runge--Kutta RK4 scheme).
Additional steps may be necessary in special cases, too.
For example,
density diffusion terms may be applied after the integration step in certain formulations
\cite{ferrand_2017}
or boundary conditions may need to be computed by extrapolating SPH-averaged values from the fluid
to the boundary \cite{adami_2012,morris_1997},
or it may be necessary to compute the apparent viscosity before the forces computation
when modelling non-Newtonian fluids.

\begin{figure}
\centerline{\includegraphics[width=\columnwidth]{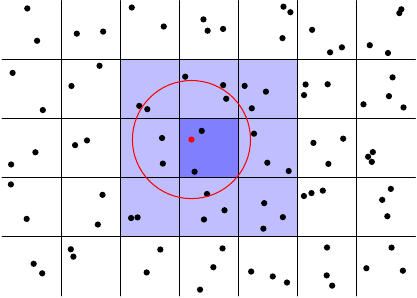}}
\vskip4ex
\centerline{\includegraphics[width=\columnwidth]{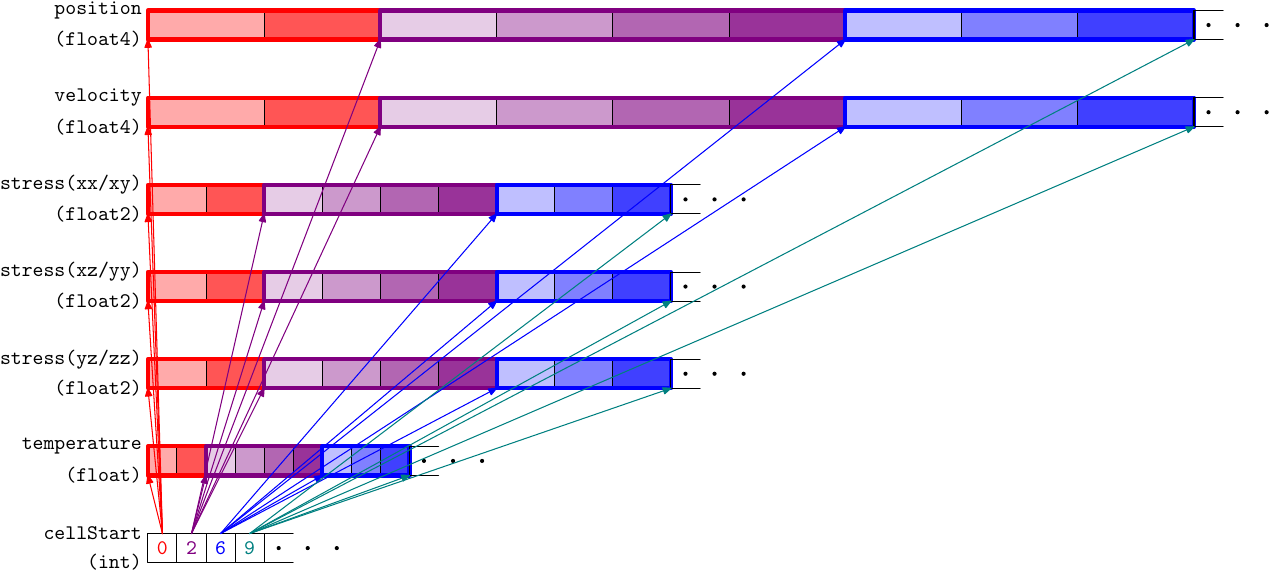}}
\caption{Grouping particles by reference cells with a cell-side which is not less that the neighbors search radius
allows the search to be limited to the particles in the Moore neighborhood of the cell to which the central particle belongs (top).
This requires a sorting process where the actual particle data is moved in memory so that the particles in the same cell
are located next to each other. An additional array can be used to track the offset in memory where the data for the particles in each cell begins (bottom).}
\label{fig:search-radius}
\end{figure}

On stream processing hardware, each of these steps will be enshrined in one or more \emph{computational kernel}s,
functions associated with the central particle, and parallelized over the entire system
according to the hardware capabilities.
In most cases, the ideal storage system for the particle properties themselves
(position, velocity, mass, density, apparent viscosity etc)
is that of a \emph{structure of arrays}, where an individual array is used for each property,
optionally merging some scalar and vector properties that are frequently used together:
for example, in GPUSPH we use a single 4-component vector data type to store 3D position and mass,
and another 4-component vector to store velocity and density.
This is especially convenient for hardware such as GPUs,
but is actually useful on most modern CPU systems as well
\cite{bilotta_design_2018}, as it tends to naturally map array elements to the hardware vector types.
Conversely,
particle data that requires more than 4 components (e.g. symmetric tensors that require 6 components in 3D)
may be inefficient to access on GPUs;
in this case it may be convenient to split the storage into smaller units,
such as a 4-component vector and a 2-component vector, or 3 2-component vectors,
as illustrated in Figure~\ref{fig:search-radius}.

\subsection{Optimizing the neighbors search}
\label{sec:nl-basis}

Each of the steps (with the possible exception of the integration steps) requires one or more loops (usually for summations) over the neighbors,
pushing the need for an efficient neighbors search.
The main strategy to improve the neighbors search performance is to adopt auxiliary data structures
that help restrict the search space.
Space partitions using trees are more common in applications where the smoothing length is variable:
for example, $n$-trees (quadtrees in two dimensions, octrees in three) to bucket particles that are close in space
are common in astrophysics, where they can be also used in support for gravity computation~\cite{gafton_2011,hernquist_1989}.
A brief review of tree-based methods, with some proposed enhancements, can be found in~\cite{cavelan}.

A straightforward auxiliary data structure that is very practical in case of fixed smoothing lengths
and that also has additional uses is a simple grid of cells with side length no less than the influence radius
(Figure~\ref{fig:search-radius}).
The grid itself is represented simply by its origin (coordinates in 2D or 3D spaces),
extents (dimensions in each coordinate direction), and grid spacing (which may be different in each coordinate direction).
If particles are sorted in memory by the cell they belong to,
bucketing can be achieved simply by storing the indices of the first and last particle in each cell.
The particle sorting also brings performance benefits related to the improved data locality~\cite{green_cuda_particles}.

The primary objective of this auxiliary grid is to limit the neighbors search to the 9 (in 2D) or 27 (in 3D) cells around the central particle cell,
but the same data structure is also useful to implement efficient multi-GPU and multi-node support \cite{rustico_multi-gpu_2014, rustico_advances_2014}
and uniform accuracy in space without the need for extended precision \cite{herault_accuracy_2014, saikali_2020},
by relying on local (cell-relative) particle positions.

Even with this improvements, the neighbors search can still be an expensive procedure,
due also to the cost of maintaining the auxiliary data structures themselves, and\slash or to the particle sort.
These costs however can be amortized (at the expense of memory consumption) by building a \emph{neighbors list} per particle,
to be rebuilt periodically~\cite{dominguez-neibs-list}.
The frequency at which the neighbors list needs to be rebuilt depends on the dynamics of the problem,
and can be reduced by using a slightly larger radius for neighbors search compared to the actual influence radius.
As we shall see in the upcoming sections, the way the neighbors list is stored, processed and built
are essential details with a significant influence on the computational performance of an SPH implementation.

\section{GPUSPH features}\label{sec:features}

\subsection{The SPH in GPUSPH}

Born with the design goal of offering a simple, high-performance implementation of classic WCSPH
\cite{herault_sph_2010},
later extended with the aim to model lava flows \cite{herault_annals_2011,bilotta_2016},
GPUSPH has since grown into a very sophisticated engine for SPH,
with the ultimate objective of becoming a \emph{universal SPH computational engine},
useful both for applications and for research in the numerical method itself.

To improve its usability in the research of WCSPH,
it is necessary for the GPUSPH code to be easily extensible,
in order to minimize the cost of implementation of new formulations,
and correct, in the sense of helping ensure that the operations are done
in the intended order and on the intended data.

For applications, it is essential that the (correct!) results are provided
in a timely manner, that they are stable (within the boundaries imposed
by the model), and that invalid data is handled appropriately.

The priorities in GPUSPH development are therefore performance and robustness.
Satisfying these goals concurrently poses a significant coding challenge,
especially in consideration of the vastness of the problem.

The SPH core of GPUSPH is the so-called ``simulation framework'', a collection
of computational kernels specialized on the basis of user-selectable options
that determine every aspect of the simulation: the equations to be solved
(e.g. Navier--Stokes, heat, both), physical aspects such as the rheological
or turbulence model, and choices about the details of the numerical model,
such as the SPH formulation, the solid boundary model, the smoothing kernel, etc.

\begin{table}
\begin{footnotesize}
\begin{tabular}{ll}
Dimensionality
& 1D, 2D, 3D \\
\hline
Smoothing kernel
& Quadratic, 
  Cubic spline, \\
& Wendland, 
  Gaussian \\
\hline
SPH formulation
& WCSPH single-fluid \cite{monaghan_2005}, \\
& WCSPH multi-fluid \cite{colagrossi_landrini_2003}, \\
& Grenier \cite{grenier_2009}, 
  Hu \& Adams \cite{hu_adams} \\
\hline
Density diffusion
& (none), \\
& Ferrari \cite{ferrari_2009,mayrhofer_2013}, 
  Brezzi \cite{Brezzi1984,ferrand_2017}, \\
& Molteni \& Colagrossi \cite{molteni_colagrossi_2009}, \\
& Antuono ($\delta$-SPH) \cite{marrone_2011}\\
\hline
Boundary model
& Lennard--Jones \cite{Liu:2003,monaghan_2005}, 
  Monaghan--Kajtar \cite{Monaghan:2009}, \\
& dynamic \cite{dalrymple2001sph,Crespo:2007}, 
  dummy \cite{adami_2012}, \\
& semi-analytical \cite{ferrand_2012,mayrhofer_2015} \\
\hline
Periodicity & (any combination of) X, Y, Z \\
\hline
Rheological models
& Inviscid, 
  Newton, \\
& Granular \cite{ghaitanellis_2021}, \\
& Bingham \cite{bingham}, 
  Papanastasiou \cite{papanastasiou}, \\
& Power-law \cite{ostwald}, \\
& Herschel--Bulkley \cite{herschel_bulkley}, 
  Alexandrou \cite{alexandrou}, \\
& DeKee \& Turcotte \cite{dekee}, 
  Zhu \cite{Zhu:2005} \\
\hline
Turbulence model
& (none), \\
& $k$-$\epsilon$ \cite{kepsilon}, 
  Sub-Particle Scale (SPS) \cite{RogersDalrympleSPH_SPS}, \\
& artificial viscosity (improperly) \cite{monaghan_2005}\\
\hline
Viscous model
& Morris \cite{morris_1997}, 
  Monaghan \cite{monaghan_2005}, \\
& Español \& Revenga \cite{espanol_revenga_2003}\\
\hline
Averaging operator
& arithmetic, 
  harmonic, 
  geometric \\
\hline
Internal viscosity
& Dynamic \\
representation
& Kinematic \\
\hline
Miscellanea
& Adaptive time-stepping \\
(boolean flags)
& XSPH \\
& Geometric planes support \\
& Geometric natural topography support \\
& Moving bodies support \\
& Open boundaries support \\
& Water depth computation \\
& Density summation \\
& Semi-analytical gamma quadrature \\
& Internal energy computation \\
& Multi-fluid support \\
& Repacking \\
& Implicit integration of the viscous term \\
\end{tabular}
\end{footnotesize}
\caption{A summary of the framework options that can be set by the user when defining a test case
in GPUSPH, with relevant bibliographical references.
Not all combinations are supported.
Some features are experimental and have not been fully merged into the public version yet.}
\label{tab:framework-options}
\end{table}

The comprehensiveness of the framework can be remarked by looking at the variety of options offered,
summarized in Table~\ref{tab:framework-options} for the latest publicly released version.
The total number of possible theoretical combinations, considering all options,
is between $10^9$ and $10^{10}$,
and that's before including additional options that have not been fully integrated in the public versions,
such as coupling with the heat equation~\cite{bilotta_2016,zago_aog_2019},
coupling with finite-element models~\cite{zago_fem_2022},
or kernel gradient corrections~\cite{zago_ccsph,schulze_ccsph_2022}.
Even though not all combinations are currently supported,
the ultimate objective remains to cover the widest possible range.
Still, to satisfy our goals of universality for both research and applications,
this must be achieved \emph{without any performance penalty for unused options},
and \emph{minimizing implementation complexity}.

The lack of performance penalty for unused options is essential for the usefulness of GPUSPH
in applications, and translates into the following maxim: the runtime of the program
when a given framework option is not used\slash enabled should be the same as if support for that
option had not been implemented in the code at all. This also means that when adding
new features, there should be no regression in the runtime of any of the pre-existing options.
This can be achieved by doing as much work as possible at compile time,
and helping the compiler in producing optimal code by isolating code and variables
that are specific to an option (see section~\ref{sec:cpp11} for details
on how this is achieved).

\subsection{Minimizing complexity}\label{sec:complexity}

\begin{figure}
\centerline{\includegraphics[width=\columnwidth]{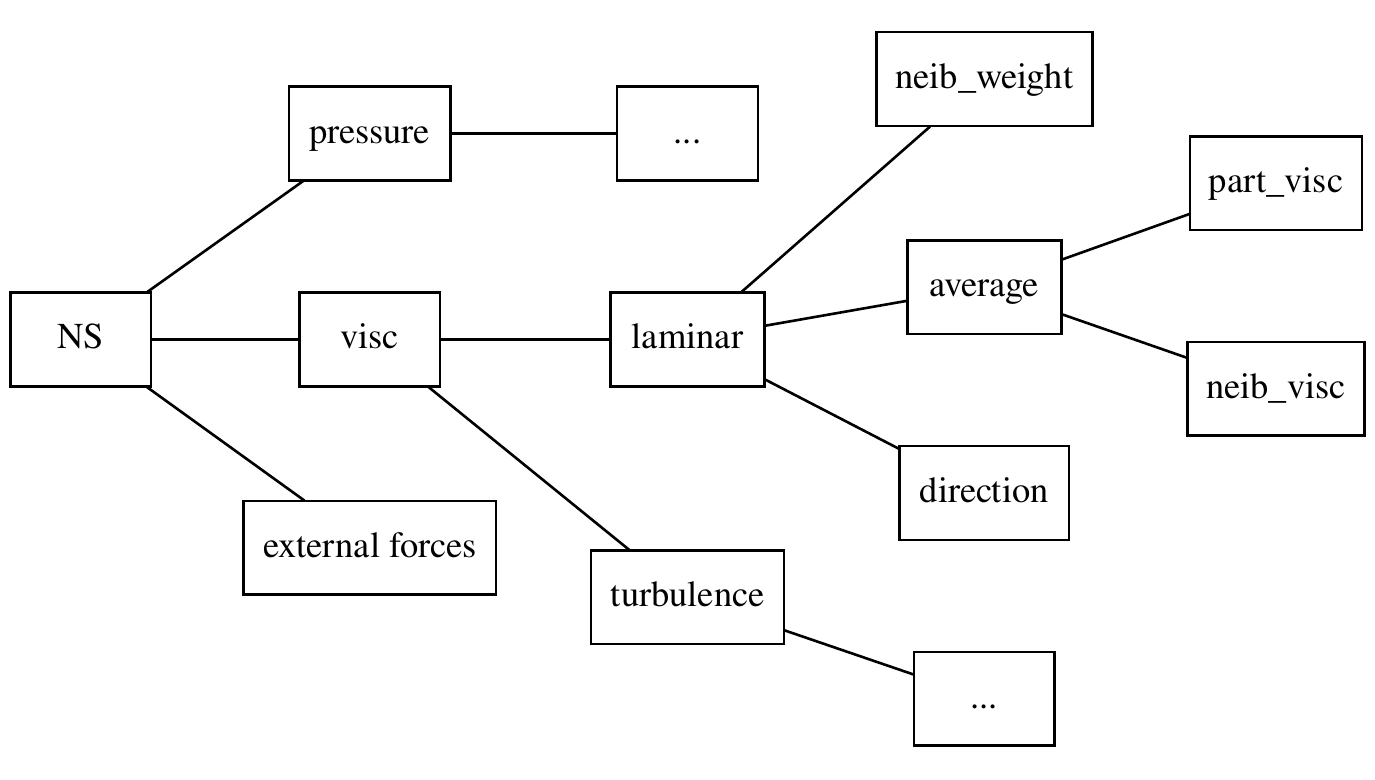}}
\caption{Call-tree for a subset of the functions involved in the computation
of particle-particle interaction for the Navier--Stokes equation.}
\label{fig:visc-forces-tree}
\end{figure}

Minimizing implementation complexity, on the other hand, is a fight against
Tesler's law of conservation of complexity~\cite{complexity}.
Our strategy is to
maintain a high level of abstraction in the code, writing small functions
and avoiding repetitions. This however comes at the cost of increased levels
of indirection (Figure~\ref{fig:visc-forces-tree}),
a more spread-out code, and the adoption of non-trivial
programming techniques
centered around the C++ idiom known as template meta-programming,
as described in Section~\ref{sec:cpp11}.

Consider for example the case of the contribution of laminar viscosity to the momentum equation
(central block in Figure~\ref{fig:visc-forces-tree}).
This can be expressed as
\[
\phantom{K}\sum_j \omega_{ij} 2 \bar\mu_{ij} F_{ij} \vec u_{ij}
\]
using Morris' formulation~\cite{morris_1997}, or as
\[
K \sum_j \omega_{ij} 2 \bar\mu_{ij} F_{ij} \frac{\vec x_{ij}\cdot \vec u_{ij}}{\vec x_{ij} \cdot \vec x_{ij}} \vec x_{ij}
\]
when using Monaghan's~\cite{monaghan_2005}, or in the form
\[
\phantom{K}\sum_j \omega_{ij} F_{ij} \left(
\left(\frac{5}{3} \bar\mu_{ij} - \bar\zeta_{ij}\right) \vec u_{ij} +
5 \left( \frac{1}{3}\bar\mu_{ij} + \bar\zeta_{ij}\right) \frac{\vec x_{ij}\cdot \vec u_{ij}}{\vec x_{ij} \cdot \vec x_{ij}} \vec x_{ij}
\right)
\]
following Español \& Revenga~\cite{espanol_revenga_2003}; in all these expressions,
$\bar\mu_{ij}$ and $\bar\zeta_{ij}$ are some average first and second viscosity,
and $\omega_{ij}$ is the volume-based neighbor weight.

The weight itself depends on the SPH formulation, being
$\omega_{ij} = \frac{m_j}{\rho_i \rho_j}$ for most formulations, but e.g.
$\omega_{ij} = \omega_j\left(\frac{1}{\sigma_i} + \frac{1}{\sigma_j}\right)$
when using Grenier's formulation (see \cite{grenier_2009} for the meaning of $\sigma$).
Likewise, the averaging operator can take many forms, including e.g. the specialized
form in the case of harmonic mean of constant \emph{kinematic} viscosity
$\bar\mu_{ij} = 2\nu \frac{\rho_i \rho_j}{\rho_i + \rho_j}$.
Finally, when possible we also aim for simplifications across terms, so that for example
$\omega_{ij} 2 \bar\mu_{ij}$ as a whole can be written as $4 \nu \frac{m_j}{\rho_i + \rho_j}$
under appropriate conditions.

This leads to specializations with complex criteria such as
\emph{single, Newtonian fluid} when using
\emph{kinematic internal viscosity representation} with \emph{harmonic averaging},
\emph{except} when using \emph{Grenier's formulation} or the \emph{Español \& Revenga viscous model}.

\begin{table*}
\begin{tabular}{l|lll}
Yield strength & \multicolumn{3}{c}{Shear rate dependency} \\
& Linear & Power-law & Exponential \\
\hline
none & Newton & Power-law & \cellcolor{gray}\\
constant & Bingham & Herschel–Bulkley & DeKee \& Turcotte \\
regularized & Papanastasiou & Alexandrou & Zhu \\
functional & Granular & \cellcolor{gray} & \cellcolor{gray} \\
\end{tabular}
\caption{Decomposition of the contributions to the rheological models supported in GPUSPH.
Separating each yield strength contribution and shear rate dependency in its own functions
allow us to support 9 (potentially $4\times3=12$) different models with only $4+3=7$
functions.}
\label{tab:rheology-factors}
\end{table*}

Finding the correct way to distribute the computations across multiple sub-functions
helps reduce the complexity of the code. For example, a set of 4 functions for
the yield strength contribution and 3 functions for the shear rate to shear stress contribution
to the apparent viscosity of generalized Newtonian fluids is sufficient to describe
9 (potentially 12) different rheological models, as summarized in Table~\ref{tab:rheology-factors}.
By separating contributions and sharing common code, we reduce the number of necessary functions
and follow the DRY (Don't Repeat Yourself) principle.

\subsection{Code structure}

GPUSPH has four main components: the already-discussed simulation framework, the integrator,
the manager and the worker(s).

Subclasses of the \code{Integrator} class describe the numerical integration scheme,
as a sequence of phases (e.g. predictor, corrector), each composed of multiple steps or commands
(e.g. apparent viscosity computation, forces computation, integration).
This allows defining different integration schemes with limited or no need to change
the other classes. GPUSPH currently implements only a predictor\slash corrector integration scheme,
and an experimental explicit Euler used only for the repacking feature.

The manager (a class named \code{GPUSPH}) represents the main thread,
and is responsible, as the name suggests, for \emph{managing} the simulation,
dispatching the commands (provided by the \code{Integrator}) to the workers or to secondary classes
(such as the writers, responsible for saving the simulation results).

The workers are secondary threads that manage the device side of the simulation.
One instance of the worker thread is created for each device (GPU),
and it takes care of tasks such as memory allocation,
cross-device data transfers, and issuing the actual computational kernels
implementing the numerical method.

The computational kernels themselves are defined by the simulation framework,
which is composed of distinct modules called ``engines'',
each dedicated to a separate aspect of the algorithm:
\begin{description}
\item[\code{neibs} engine] handles the construction of the neighbors list,
including the reordering of the particle data so that particles close in space
are also close in device memory (see Section~\ref{sec:nl-basis});
\item[\code{visc} engine] handles the computations related to non-Newtonian
and turbulent viscous models;
\item[\code{boundary} engine] handles the computations related to imposing
boundary conditions;
\item[\code{forces} engine] handles the computation of the ``forces''
of the particles, in the generalized sense of time derivatives of their state variables
(including e.g. density and temperature);
\item[\code{euler} engine] handles integration in time
(despite the name of the engine, GPUSPH doesn't actually use the forward Euler method directly,
but an explicit predictor\slash corrector integration scheme,
with a forthcoming option for a semi-implicit scheme \cite{zago_jcp_2018,bilotta_bicgstab}).
\end{description}

Each engine is described by an abstract interface, and implemented
by classes that specialize the interface based on the framework options.
This allows the engine methods to be invoked by the workers 
with minimal knowledge about the framework options being used,
delegating the gory details of the exact nature of the computations that
need to be executed to the \emph{computational kernels} defined by the specific engine incarnation.

This separation of roles is essential both for the multi-GPU and multi-node support in GPUSPH,
and to reduce the complexity of implementation of new features: the latter effect
is due to the reduction of the influence of the features onto each component
to a ``need-to-know'' basis.


\begin{lstfloat}[caption={Subset of the \code{AbstractForcesEngine} definition, showing the interface
for the \code{forces} engine used to compute particle dynamics.},
label={lst:AbstracForcesEngine}]
class AbstractForcesEngine {
public:
	/* omissis */

	/// Compute density diffusion term
	virtual void
	compute_density_diffusion(
		const BufferList& bufread,
		BufferList& bufwrite,
		const	uint	numParticles,
		const	uint	particleRangeEnd,
		const	float	deltap,
		const	float	slength,
		const	float	influenceRadius,
		const	float	dt) = 0;

	/* omissis */
}
\end{lstfloat}

\begin{lstfloat}[caption={Implementation of a command in the worker class,
mapping the symbolic command name to the corresponding \code{AbstractForces} member function.
Most command implementations follow the same structure.},
label={lst:runCommandExample}]
template<>
void GPUWorker::runCommand<CALC_DENSITY_DIFFUSION>(CommandStruct const& cmd)
{
	uint numPartsToElaborate = (cmd.only_internal ? m_particleRangeEnd : m_numParticles);

	if (numPartsToElaborate == 0) return;

	const int step = cmd.step.number;

	const BufferList bufread = extractExistingBufferList(m_dBuffers, cmd.reads);
	BufferList bufwrite = extractExistingBufferList(m_dBuffers, cmd.updates) |
		extractGeneralBufferList(m_dBuffers, cmd.writes);

	bufwrite.add_manipulator_on_write("calcDensityDiffusion" + to_string(step));

	const float dt = cmd.dt(gdata);

	forcesEngine->compute_density_diffusion(
		bufread,
		bufwrite,
		m_numParticles,
		numPartsToElaborate,
		gdata->problem->m_deltap,
		m_simparams->slength,
		m_simparams->influenceRadius,
		dt);

	bufwrite.clear_pending_state();
}
\end{lstfloat}

Examples of the interfaces for the engines and how they are used by the workers are shown in Listings~\ref{lst:AbstracForcesEngine}
and~\ref{lst:runCommandExample}.
The command implementation is nearly the same for most commands. The \code{CommandStruct} structure, defined by the integrator,
holds metadata information about the command,
such as which particle data should be made available to the engine function, how it should be accessed
(for reading, for writing ignoring previous content, or for updating), and which value of the time-step should be used
(depending on the integration step, this could be $\Delta t$ or $\Delta t/2$).
This information is used by the worker function to extract the relevant subsets of the global list of particle data buffers,
which are then passed to the engine function, together with the number of particles (total, and to be processed),
geometric information such as the inter-particle spacing, smoothing length and influence radius, and the effective time-step being used.
Additionally, buffers that will be written to by the engine function are marked with information that
is useful when debugging, such as the commands and time-steps at which the buffer contents were modified.

Note that the worker function itself has no specific information about which buffers are going to be used for this specific command:
this information defined on the integrator side, and then used on the engine side.
This allows, for example, a new density diffusion formulation that uses different sets of particle data to be introduced by changing only
the integrator and the engine, but not the worker, that is simply passing the information along.

\subsection{The GPU in GPUSPH}

GPUSPH was designed from the ground up to rely exclusively on GPUs for the computational part~\cite{herault_sph_2010}.
To achieve this, the program execution is split into three phases: initialization, simulation, and data storage.

The initialization phase runs on program startup, on the host CPU, and it takes care of generating the initial particle distribution,
either from a geometric description of the domain or from data stored on disk (e.g. when resuming an interrupted simulation).

After initialization, the worker threads are created, instantiating a \code{GPUWorker} for each GPU selected for the simulation,
and the domain particles are distributed to the GPUs (i.e. the first and last particle assigned to each GPU is computed).
Each worker then allocates data arrays on its GPU, forming the global buffer lists,
and the buffer contents are initialized by copying data over from the host, for
the subset of particles assigned to the specific device.

These data arrays are the ones used by the computational kernels, extracted from the global buffer lists as shown in Listing~\ref{lst:runCommandExample},
and passed to the appropriate engine functions, that take care of launching the actual computational kernels on the GPU.

No further data exchange between the GPU and the host happens,
except for the following circumstances:
\begin{itemize}
\item after the particle sorting and neighbors list construction, the host fetches information about
the current number of particles, and the maximum number of neighbors per particle;
this information is used to check if particles have been removed because they had gone out of bounds,
to track the new distribution of particles in multi-GPU when particles cross from one device to an adjacent one,
and to check that all neighbors could be accounted for
(issues with the number of neighbors typically indicate either an incorrect initial particle distribution,
or some issues with the choice of formulation or its implementation);
\item the maximum allowed time-step (minimum over all the particles) is computed on device using a parallel reduction,
and then downloaded to the host, for time-keeping;
\item for fluid\slash structure interaction, the cumulative forces and torques exerted by the fluid on each rigid body
are computed on the GPU, and downloaded to the host, to be passed to Project Chrono to compute the motion of the rigid body;
the updated position of the center of mass and rotation of the rigid bodies is then copied back to the GPU,
where the information is used to move the body particles accordingly.
\end{itemize}

For multi-GPU (both single- and multi-node) simulations,
data is transferred directly from device to device if possible, i.e. if the GPUs can access each other's memory
either through peering (on one machine) or through GPUDirect in multi-node configurations where the network setup supports it~\cite{rustico_advances_2014}.
When this is not possible, data transfer happens through a staging area on host, which can negatively affect performance
due to the additional memory copies.

Multi-GPU data transfers are explicitly marked by the integrator,
allowing the developer to choose when to transfer data, and which data to transfer.
These choices can be tuned to improve scaling by overlapping computations and data transfer~\cite{rustico_multi-gpu_2014,rustico_advances_2014}.

Finally, to allow data storage to disk, all particle data arrays get downloaded to the host
at fixed (simulated) time interval selected by the user.
The simulation is suspended during this process.

It should be noted also that the arrays are only allocated once during the initialization phase.
If the number of particles decreases during the simulation, e.g. because some particles fly out of the computational domain,
the contents of the arrays are compacted during sorting and the additional entries are simply ignored.
Allocations are made taking into account the possibility of the number of particles increasing becase of open boundaries or in the multi-GPU case;
in the open boundary case the user has control on the maximum number of particles that may be considered for the simulation,
and in case of overflow the program terminates, allowing the user to resume with a higher upper bound.
This avoids expensive reallocations at runtime.

\section{C++11 and the benefits of template meta-programming}\label{sec:cpp11}

The abstract interfaces of the GPUSPH engines is preserved down the stack whenever possible.
In particular, many computational kernels are as generic as possible
as well, with the details further delegated to auxiliary functions, and so forth
down to small, specialized functions that materialize one specific aspect
of the computation, as exemplified in section~\ref{sec:complexity}.
The key to writing such generic code in C++ is an extensive use of \emph{template}s,
combined with the coding strategies described in the following subsections,
as illustrated in Listings~\ref{lst:forces-methods} and~\ref{lst:forces-kernel} below.

A function template represents a family of functions that share a name,
expose the same interface, and also frequently (but not necessarily) present the same body.
Such a family of functions is parametrized through one or more \emph{template parameter}s,
with each possible value (or combination of values) of these parameters ultimately dictating
the actual function.

Template parameters are usually types, as in the classic examples
of the \code{std::min} and \code{std::max} function templates
from the C++ standard library, that allow the same function body
(e.g. \code{return a < b ? a : b}) to be instantiated for
each data type (\code{int}, \code{float}, \code{double}) without the known issues
of the C preprocessor macros.

In GPUSPH however, most template parameters are enumeration types,
such as the symbolic name of the SPH formulation or of the boundary conditions model.
For example, the \code{W} and \code{F} function templates,
used to compute the value of the smoothing kernel and the scalar part of its gradient,
only depend on one template parameter, the \code{KernelType} enumeration value identifying the
user choice of smoothing kernel.
In this case, each specialization of the function templates \code{W} and \code{F}
has a different body, since it must implement different computations
(Listing~\ref{l:ktemplate}).

\begin{lstfloat}%
[caption={Function templates for the smoothing kernel $W(\cdot, h)$ and its scaled gradient $F(\cdot, h)$.
Some specialized instance are presented for demonstrative purposes.
The normalization coefficients and the Gaussian truncation factor are device constants that are precomputed
at the beginning of the simulation.},
 label={l:ktemplate}]
enum KernelType {
	CUBICSPLINE, QUADRATIC, WENDLAND, GAUSSIAN,
};

template<KernelType kerneltype>
float W(const float r, const float slength);
template<KernelType kerneltype>
float F(const float r, const float slength);

// Wendland kernel: w(q) = (2q+1)(1 - q/2)^4, f(q) = (q-2)^3
template<>
float W<WENDLAND>(const float r, const float slength) {
	const float R = r/slength;

	float val = 1.0f - 0.5f*R;
	val *= val; val *= val;
	val *= 1.0f + 2.0f*R;
	val *= d_wcoeff_wendland;
	return val;
}
// Wendland kernel: f(q) = (q-2)^3
template<>
float F<WENDLAND>(const float r, const float slength)
{
	const float qm2 = r/slength - 2.0f;
	float val = qm2*qm2*qm2*d_fcoeff_wendland;
	return val;
}

// d-truncated Gaussian kernel: w(q) = exp(-q^2) - exp(-(d/h)^2)
template<>
float
W<GAUSSIAN>(float r, float slength)
{
	const float R = r/slength;

	float val = expf(-R*R);
	val -= d_wsub_gaussian;
	val *= d_wcoeff_gaussian;
	return val;
}
// Truncated Gaussian kernel: f(q) = -exp(-q^2)
template<>
float F<GAUSSIAN>(const float r, const float slength)
{
	const float R = r/slength;
	float val = -expf(-R*R)*d_fcoeff_gaussian;
	return val;
}

/* and so on for the other smoothing kernels */
\end{lstfloat}


The advantage of function templates is that they eliminate a subset of the runtime conditionals,
maximizing performance.
For example, instead of checking which smoothing kernel was requested by the user
every time \code{W} or \code{F} is evaluated, the selection is made once at compile time,
eliminating conditionals from the hot path.

The downside is that the template parameters must be known at compile-time,
since they are needed by the compiler to determine which versions of the function to emit,
depending on what is being called.
Because of this, template arguments ``creep'' up in the call chain up to the point where the user choice is given,
affecting all functions and classes that directly or indirectly result in these functions being called.
For example, any function (or computational kernel) that needs to call the \code{W} or \code{F} function templates
will have to be a function template depending on the \code{KernelType} enumeration,
and thus the framework engine classes that need to call these computational kernels
will also have to be class templates that depend the same template parameter.
Listings~\ref{lst:forces-methods} and~\ref{lst:forces-kernel} show an example of this.

\begin{lstfloat}[caption={Declaration of the \code{CUDAForcesEngine} class template,
concrete class implementing the \code{AbstractForcesEngine} interface, and example
of a member function called by the appropriate \code{GPUWorker::runCommand()} specialization.},
label={lst:forces-methods}]
template<
	KernelType kerneltype,
	SPHFormulation sph_formulation,
	DensityDiffusionType densitydiffusiontype,
	typename ViscSpec,
	BoundaryType boundarytype,
	flag_t simflags>
class CUDAForcesEngine : public AbstractForcesEngine
{

/* omissis */

void
compute_density_diffusion(
	BufferList const& bufread,
	BufferList& bufwrite,
	/* omissis */)
{
	uint numThreads = BLOCK_SIZE_FORCES;
	uint numBlocks = div_up(particleRangeEnd, numThreads);

	auto params = density_diffusion_params<kerneltype, sph_formulation, densitydiffusiontype,
		boundarytype, PT_FLUID>
		(bufread, bufwrite, /* omissis */);

	cuforces::computeDensityDiffusionDevice
		<kerneltype, sph_formulation, densitydiffusiontype, boundarytype,
		 ViscSpec, simflags, PT_FLUID>
		<<<numBlocks, numThreads>>>(params);

	// check if last kernel invocation generated an error
	KERNEL_CHECK_ERROR;
}

/* omissis */

};
\end{lstfloat}

\begin{lstfloat}[caption={Declaration of the computational kernel called by the corresponding member function
of \code{CUDAForcesEngine}, implementing the actual computations, and its arguments structure template,
leveraging conditional subclassing to determine which data members (i.e. function arguments)
are present.},
label={lst:forces-kernel}]
struct common_density_diffusion_params
{
			float4	* __restrict__ forces;
	const	float4	* __restrict__ posArray;
	const	float4	* __restrict__ velArray;
	/* omissis */

	common_density_diffusion_params(
		BufferList const&	bufread,
		BufferList &		bufwrite,
		/* omissis */)
	:
		forces(bufwrite.getData<BUFFER_FORCES>()),
		posArray(bufread.getData<BUFFER_POS>()),
		velArray(bufread.getData<BUFFER_VEL>()),
		/* omissis */
	{}
};

template<
	KernelType _kerneltype,
	SPHFormulation _sph_formulation,
	DensityDiffusionType _densitydiffusiontype,
	BoundaryType _boundarytype,
	ParticleType _cptype>
struct density_diffusion_params :
	common_density_diffusion_params,
	cond_struct<_boundarytype == SA_BOUNDARY, sa_boundary_params>
{
	/* omissis */

	density_diffusion_params(
		BufferList const&	bufread,
		BufferList &		bufwrite,
		/* omissis */)
	:
		common_density_diffusion_params(bufread, bufwrite, /* omissis */),
		cond_struct<_boundarytype == SA_BOUNDARY, sa_boundary_params>(bufread)
	{}
};

template<KernelType kerneltype,
	SPHFormulation sph_formulation,
	DensityDiffusionType densitydiffusiontype,
	BoundaryType boundarytype,
	typename ViscSpec,
	flag_t simflags,
	ParticleType cptype>
__global__ void
computeDensityDiffusionDevice(density_diffusion_params<kerneltype, sph_formulation,
	densitydiffusiontype, boundarytype, cptype> params)
{
	// Global particle index
	const uint index = INTMUL(blockIdx.x,blockDim.x) + threadIdx.x;

	if (index >= params.particleRangeEnd) return;

	/* omissis: actual computation of the density diffusion contribution */
}
\end{lstfloat}

A class template represents a family of classes that expose the same (or similar)
interfaces, member variables and member functions (methods). Classic examples
in the C++ standard library are the container classes such as \code{std::vector}
or \code{std::map}. In GPUSPH, class templates are used extensively in the
host code to represent things such as buffers, integrator schemes,
the simulation framework and its engines,
and in device code for the variables and arguments structures
that will be discussed in the following sections.

A powerful use of class or structure templates in C++ is the \emph{traits}
concept: these are structure templates that hold a collection
of properties associated with specific values of its template parameters.
In the standard library, for example, the representation limits of
each data type (minimum and maximum representable value, number of digits etc)
are collected into the \code{std::numeric\_limits} structure template
(rather than in preprocessor macros as in the C standard).

One of the primary uses of type traits in GPUSPH is the \code{BufferTraits} structure template,
that associates the symbolic name of each buffer with the type of the elements,
the number of arrays associated with the buffer,
and a user-readable name.
For example, the traits declare \code{BUFFER\_POS} to be a single array of \code{float4}
elements, named ``Position'', while \code{BUFFER\_TAU} is a triple array of \code{float2}
elements (needed to store the 6 elements of a symmetric tensor) named ``Tau''.
This information allows GPUSPH to do type-checking on the buffers when they get passed
to computational kernels, and to support most memory management and data exchange
(between host and devices on initialization and when saving data,
or between devices in multi-GPU and multi-node setups)
automatically through common interfaces.

Type traits are just one of the keys to \emph{template meta-programming}~\cite{moderncppdesign},
a~C++ coding strategy that improves code abstraction and minimization.
With version~5 of GPUSPH, and the drop of support for older version of CUDA that
had so far prevented us from adopting more recent versions of the C++ standard,
we have finally transitioned to C++11~\cite{cpp11} as the minimum C++ version
for both the host and the device code. This has allowed us to
leverage the facilities offered by the new language revision
to improve the compactness and legibility (and sometimes the expressive power)
of our templates.

We will discuss here three meta-programming techniques heavily used in the device code of GPUSPH,
to show how they allow us to minimize repetition (DRY principle)
while also reducing the compiler-visible variables
and function arguments to the minimum necessary for each specialization of
a computational kernel or function. All of these features
can be implemented without relying on C++11 features,
but the syntax on older revisions of the C++ language is considerably
more convoluted to use.
Much of the syntax could be further simplified by raising the minimum requirements of
more recent versions of CUDA that support C++14, but we chose to compromise to extend
the range of hardware supported by GPUSPH, since newer CUDA versions tend to deprecate
support for older hardware.

\subsection{Conditional structures}
\label{sec:cond-struct}

One of the main complexities associated with supporting different physical-numerical models
is that each of them will require keeping track of different state variables
for the particles, and will need different function arguments to be passed
to the computational kernels and their auxiliary functions.

For example, when using the $k$-$\epsilon$ turbulence model
\cite{kepsilon}
one will need to track $k, \epsilon$ and their derivatives,
whereas these properties (and all the associated computations) will not be needed
when using a different (or no) turbulence model.
Likewise, when using the ``dummy boundary'' \cite{adami_2012} model,
boundary particles will have an associated ``viscous velocity'' used
when computing the viscous term in the momentum equation for neighboring
fluid particles, but the particle velocity alone may suffice for all other
particles and/or other boundary models.

To help the compiler produce optimal code, and to avoid unintended mangling
due to developer error, we wish the corresponding variables and function arguments
to not be \emph{present} in the computational kernel specializations
that do not need them.

Focusing on variables (the discussion is exactly the same for arguments),
the basic idea is that, instead of declaring each variable
independently, we define a structure whose members correspond to the original variables.
This transforms the problem of not having unnecessary variables
to the problem of not having unnecessary members in the structure.

The transformed problem can be solved using the \emph{conditional inheritance} paradigm,
i.e. the possibility to have a class templates
declare different base classes depending on the template parameters.
The idea is to collect optional members (variables) into their own structures,
and then have the full structure derive only from the substructures
that are required by the specific combination of framework options.
This can be achieved using the \code{std::conditional}
structure template provided from C++11 onwards,
which is defined in such a way that
\begin{lstdisplay}
std::conditional<b, TrueClass, FalseClass>::type
\end{lstdisplay}
will be equivalent to \code{TrueClass} if the boolean value \code{b} is true,
and to \code{FalseClass} if it's false.
On older revisions of C++, the same structure template can be defined by the user
in the following way:
\begin{lstdisplay}
template<bool B, typename T, typename F>
struct conditional { typedef T type; };

template<typename T, typename F>
struct conditional<false, T, F> { typedef F type; };
\end{lstdisplay}
although some of the syntax simplifications illustrated later on will not be possible.

By making the \code{FalseClass}
an empty structure, we can achieve our goal of selective inclusion of member variables.
Given for example the structure template presented in Listing~\ref{l:simple-empty},
the \code{all\_variables<DUMMY\_BOUNDARY>} specialization would have
members \code{dummy\_vel}, \code{pos} and \code{vel}, whereas
a specialization for a different \code{BoundaryType} (e.g. \code{all\_variables<DYN\_BOUNDARY>}) would only
have \code{pos} and \code{vel}, since the base structure would
be the memberless \code{empty} structure in this case.

\begin{lstfloat}%
[caption={Simple application of conditional inheritance.
The \code{all\_variables} structure template will only have the \code{dummy\_vel}
member in the \code{all\_variables<DUMMY\_BOUNDARY>} specialization.},
 label=l:simple-empty]
struct empty {};

struct dummy_boundary_variables {
	float3 dummy_vel;
};

template<BoundaryType boundarytype>
struct all_variables :
	std::conditional<boundarytype == DUMMY_BOUNDARY,
		dummy_boundary_variables, empty>::type
{
	float3 pos;
	float3 vel;
};
\end{lstfloat}

There are two issues with this naive approach: one is
that we cannot have multiple independent conditional base structures,
and the other is related to constructors for the optional classes.

For the case of multiple optional components,
consider the example in Listing~\ref{l:double-empty-fail}
where we have added an additional optional structure holding the variables
for the $k$-$\epsilon$ turbulence model. Now, when neither
of the optional base structures are needed
(e.g. for an \code{all\_variables<DYN\_BOUNDARY, SPS>} combination),
the specialization of \code{all\_variables} would
have the \code{empty} structure as base class \emph{twice},
which is not allowed in~C++.

\begin{lstfloat}%
[caption={Non-working example of conditional inheritance with multiple optional components.},
 label=l:double-empty-fail]
struct empty {};

struct dummy_boundary_variables {
	float3 dummy_vel;
};

struct ke_variables {
	float turb_k;
	float turb_eps;
};

template<BoundaryType boundarytype, TurbulenceModel turbmodel>
struct all_variables :
	std::conditional<boundarytype == DUMMY_BOUNDARY,
		dummy_boundary_variables, empty>::type
	std::conditional<turbmodel == KEPSILON,
		ke_variables, empty>::type
{
	float3 pos;
	float3 vel;
};
\end{lstfloat}

The solution is to turn \code{empty} itself in a structure \emph{template},
whose argument is the class that is going to be replaced:
this will make each specialization formally unique (Listing~\ref{l:double-empty-correct}),
although it makes the syntax considerably heavier.

\begin{lstfloat}%
[caption={Working example of conditional inheritance with multiple optional components.},
 label=l:double-empty-correct]
template<typename T>
struct empty {};

struct dummy_boundary_variables {
	float3 dummy_vel;
};

struct ke_variables {
	float turb_k;
	float turb_eps;
};

template<BoundaryType boundarytype, TurbulenceModel turbmodel>
struct all_variables :
	std::conditional<boundarytype == DUMMY_BOUNDARY,
		dummy_boundary_variables,
		empty<dummy_boundary_variables> >::type,
	std::conditional<turbmodel == KEPSILON,
		ke_variables, empty<ke_variables> >::type
{
	float3 pos;
	float3 vel;
};
\end{lstfloat}

With C++11, the syntax can be simplified by using \emph{type alias templates}
like in Listing~\ref{l:type-alias},
whereas with older C++ versions a less robust preprocessor macro can be used
for the same effect.

\begin{lstfloat}%
[caption={Type alias templates to simplify the optional base class syntax.},
 label=l:type-alias]
template<bool B, typename T>
using cond_struct = typename std::conditional<B, T, empty<T>>::type;

template<BoundaryType boundarytype, TurbulenceModel turbmodel>
struct all_variables :
	cond_struct<boundarytype == DUMMY_BOUNDARY,
		dummy_boundary_variables>,
	cond_struct<turbmodel == KEPSILON, ke_variables>
{
	float3 pos;
	float3 vel;
};
\end{lstfloat}

The remaining issue pertains the construction of the base classes.
We would like a syntax such as the one in Listing~\ref{l:optional-construct}
to work for both the true and false cases, but this requires
the \code{empty} template to accept the same arguments
as the structures it replaces, which may be varied in number and type.

\begin{lstfloat}%
[caption={Constructing the variables structure from the constructors of the optional base classes.},
 label=l:optional-construct]
template<BoundaryType boundarytype, TurbulenceModel turbmodel>
struct all_variables :
	cond_struct<boundarytype == DUMMY_BOUNDARY,
		dummy_boundary_variables>,
	cond_struct<turbmodel == KEPSILON, ke_variables>
{
	float3 pos;
	float3 vel;

	all_variables(/* arguments used to construct all members,
		including the optional ones */)
	:
		cond_struct<boundarytype == DUMMY_BOUNDARY,
				dummy_boundary_variables>
			(/* arguments to construct dummy_boundary_variables */),
		cond_struct<turbmodel == KEPSILON, ke_variables>
			(/* arguments to construct ke_variables */),
		pos(/* arguments to construct pos */),
		vel(/* arguments to construct vel */)
	{}
};
\end{lstfloat}

The solution is to provide the \code{empty} template with a generic
constructor that can take any argument (and ignore it).
This could be simply a variadic constructor \code{empty(\dots) \{\}},
but this is not supported in CUDA code by some compiler version,
even when the effect of the constructor is trivial like in the case of the \code{empty} template.

The alternative, assuming C++11 support, is to use variadic templates for the
same purpose (the universal constructor in Listing~\ref{l:variadic-empty}).
With older C++ revisions, the only solution is to provide one template constructor for each given number of arguments,
as shown in Listing~\ref{l:variadic-empty} for the alternative preprocessor branch.

\begin{lstfloat}%
[caption={The full \code{empty} template, with constructors.},
 label=l:variadic-empty]
template<typename>
struct empty
{
	// Empty constructor
	empty() {}

#if __cplusplus >= 201103L
	// C++11 and higher: universal empty constructor
	template<typename ...T1>
	empty(T1 const&...) {}

#esle // __cplusplus < 201103L
	// Before C++11: one constructor per number of arguments
	template<typename T1>
	empty(T1 const&) {}

	template<typename T1, typename T2>
	empty(T1 const&, T2 const&) {}

	template<typename T1, typename T2, typename T3>
	empty(T1 const&, T2 const&, T3 const&) {}

	// .., and so on as needed
#endif
};
\end{lstfloat}

In GPUSPH, this technique was first used for the \code{forces} computational kernel,
which is the most complex and computationally intensive kernel, for which
we defined a \code{forces\_params} structure template to collect
all the arguments (mostly: arrays of particle data) to be passed to
the \code{forces} specializations, and for the variables structures,
collected by use: \code{particle\_data} to collect the properties of the current particle,
\code{neib\_data} to collect the properties of the current neighbor,
\code{neib\_output} to collect the contributions from the current neighbor,
and \code{particle\_output} to gather all of the contributions
affecting the evolution of the current particle (Listing~\ref{l:forcestemplates}).

\begin{lstfloat}%
[caption={The declarations of the parameters structure and variables structures used by the forces computation kernel.},
 label=l:forcestemplates]
template<KernelType kerneltype,
	SPHFormulation sph_formulation,
	BoundaryType boundarytype,
	ViscosityType visctype,
	flag_t simflags>
struct forces_params;

template<KernelType kerneltype,
	SPHFormulation sph_formulation,
	BoundaryType boundarytype,
	ViscosityType visctype,
	flag_t simflags>
struct particle_data;

template<KernelType kerneltype,
	SPHFormulation sph_formulation,
	BoundaryType boundarytype,
	ViscosityType visctype,
	flag_t simflags>
struct neib_data;

template<KernelType kerneltype,
	SPHFormulation sph_formulation,
	BoundaryType boundarytype,
	ViscosityType visctype,
	flag_t simflags>
struct neib_output;

template<KernelType kerneltype,
	SPHFormulation sph_formulation,
	BoundaryType boundarytype,
	ViscosityType visctype,
	flag_t simflags>
struct particle_output;
\end{lstfloat}

Historical note: adopting this strategy in GPUSPH did not produce any performance benefit,
since it replaced an ad-hoc solution where a similar effect was obtained
through an approach where the same “model” body of the kernel
was defined in its own file that was included multiple times,
each with a different set of defines that would identify the specific
variant of the kernel to be built, and using conditional preprocessing
to exclude unnecessary variables, following what is known as the X-Macro technique.
This strategy was actually imposed on us by limitations in the number of template parameters
in first versions of the CUDA compilers.

Even without performance benefits,
conditional inheritance still presents significant advantages compared to the X-Macro approach,
particularly in terms of source code simplification, higher readability with no preprocessor conditionals,
and most importantly a significantly reduced developer burden when adding new features,
and particularly new template parameters.
The latter benefit was additionally boosted by its combination with the thorough refactoring
allowed by the specialization based on the
SFINAE (Specialization Failure Is Not An Error) principle,
that will be described further on (section~\ref{sec:sfinae}).

\subsection{Reducing the number of template arguments}
\label{sec:fp}

The extensive use of arguments and variables structures propagates
to all auxiliary function templates. In fact, most if not all variables structures
in the computational kernels (or rather kernel templates) are effectively also
the arguments structures of the auxiliary functions (or rather function templates)
called by the kernel. This can cause an increase in the maintenance burden.

Indeed, as the number of framework options grows, so does the list of
template arguments to computational kernel templates
and their arguments and variables structure templates,
as well as to the auxiliary function templates to which the computations are delegated.
This inflicts a high development cost, since the introduction of a
new set of framework options requires changing the list of template parameters
of most functions, even when many of them may not be actually affected by the new parameter
(in the sense that its value wouldn't alter the function definition or behavior),
and the only reason why the function needs to have the template parameter is to be able to define
the type of its arguments structure.

\begin{lstfloat}%
[caption={Propagating individual framework options as template arguments.},
 label=l:templatelots]
template<KernelType kerneltype,
	SPHFormulation sph_formulation,
	BoundaryType boundarytype,
	ViscosityType visctype,
	flag_t simflags>
void particleParticleInteraction(
	particle_data<kerneltype, boundarytype, visctype, simflags>
		const& pdata,
	neib_data<kerneltype, boundarytype, visctype, simflags>
		const& ndata,
	neib_output<kerneltype, boundarytype, visctype, simflags>&
		nout,
	particle_output<kerneltype, boundarytype, visctype, simflags>&
		pout)
{ /* omissis: compute particle-particle interaction */ }

template<KernelType kerneltype,
	SPHFormulation sph_formulation,
	BoundaryType boundarytype,
	ViscosityType visctype,
	flag_t simflags>
void forces(
	forces_params<kerneltype, boundarytype, visctype, simflags> params)
{
	particle_data<kerneltype, boundarytype, visctype, simflags>
		pdata( /* omissis: initialize particle data */ )
	particle_output<kerneltype, boundarytype, visctype, simflags>
		pout;

	for_each_neighbor( /* omissis */ ) {
		neib_data<kerneltype, boundarytype, visctype, simflags>
			ndata( /* omissis: initialize particle data */ )
		neib_output<kerneltype, boundarytype, visctype, simflags>
			nout;
		particleParticleInteraction(pdata, ndata, nout, pout);
	}

	/* omissis */
}
\end{lstfloat}

This is illustrated in Listing~\ref{l:templatelots}: adding a new framework option
(e.g. to enable support for the heat equation)
would require it to be included not only in the definition of all the arguments and variables
structures seen in Listing~\ref{l:forcestemplates},
that would need it to determine which new members to include through the technique described in Section~\ref{sec:cond-struct},
but also in the declaration of \code{forces}, \code{particleParticleInteraction}
and any functions called by them and taking those same structures as input.

The solution we have adopted in GPUSPH is to use a single template parameter for
each argument: the unspecified type of the argument itself.
Referring to the same example in Listing~\ref{l:templatelots},
the singatures for \code{forces} and \code{particleParticleInteraction} would simplify to
\begin{lstdisplay}
template<typename FP> void forces(FP params);

template<typename P, typename N, typename NO, typename PO)
void particleParticleInteraction
	(P const& pdata, N const& ndata, NO& nout, PO& pout);
\end{lstdisplay}
where \code{FP} is assumed to be
some specialization of the \code{forces\_params} arguments structure template,
and the typenames \code{P, N, NO, PO}
are assume to stand for some specialization of
\code{particle\_data}, \code{neib\_data}, \code{neib\_output}
and \code{particle\_output} respectively.

When the body of the function template doesn't depend on the specific specialization of the parameter structure template,
this allows us to pass on the differentiation down to any called function that needs it
When differentiation is wanted, it will be necessary to access the original template parameters
(i.e. the values of the framework options):
e.g. we might want to know which specific value of \code{BoundaryType} was used to specialize
the \code{forces\_params} structure that is passed to the kernel.
C++ does not provide such a feature, but it can be
achieved by adding appropriate static members to the structure template:
a \code{static constexpr} member for each value type template parameter,
and a type alias for each typename template parameter
(\code{constexpr}, introduced in C++11, can be replaced with a simple \code{const} on older revisions of C++).

\begin{lstfloat}%
[caption={Static constant members can be used to expose the template parameters used to specialize the structure.},
 label=l:constexpr]
template<KernelType kerneltype_,
	SPHFormulation sph_formulation_,
	BoundaryType boundarytype_,
	ViscosityType visctype_,
	flag_t simflags_>
struct forces_params /* omissis: conditional base classes */
{
	static constexpr KernelType kerneltype = kerneltype_;
	static constexpr SPHFormulation sph_formulation =
		sph_formulation_;
	static constexpr BoundaryType boundarytype =
		boundarytype_;
	static constexpr ViscosityType visctype = visctype_;
	static constexpr flag_t simflags = simflags_;

	/* omissis: constructors etc */
};

template<typename P,  /* specialization of particle_data */
	typename N,  /* specialization of neib_data */
	typename NO, /* specialization of neib_output */
	typename PO> /* specialization of particle_output */
void particleParticleInteraction(Pt const& pdata, Nt const& ndata,
	NO& nout, PO& pout)
{ /* omissis: compute particle-particle interaction */ }

template<FP /* specialization of forces_params */>
void forces(FP const& params)
{
	particle_data<FP::kerneltype, FP::boundarytype,
		FP::visctype, FP::simflags> pdata
			( /* omissis: initialize particle data */ )
	particle_output<FP::kerneltype, FP::boundarytype,
		FP::visctype, FP::simflags> pout;

	for_each_neighbor( /* omissis */ ) {
		neib_data<FP::kerneltype, FP::boundarytype,
			FP::visctype, FP::simflags> ndata
				( /* omissis: initialize particle data */ )
		neib_output<FP::kerneltype, FP::boundarytype,
			FP::visctype, FP::simflags> nout;

		particleParticleInteraction(pdata, ndata, nout, pout);
	}

	/* omissis */
}
\end{lstfloat}

This is illustrated in Listing~\ref{l:constexpr},
where the specific values for the kernel type, boundary model, etc
used to defined the specialization of \code{forces\_params}
that will be passed to \code{forces} are accessible within the function
itself through \code{FP::kerneltype}, \code{FP::boundarytype}, etc.
Moreover, this information will be available at compile time, exactly as if
it had been specified as a template parameter to the function itself,
and can thus be used e.g. in the specification of additional dependent types
or auxiliary functions, which is used in Listing~\ref{l:constexpr}
to specify the template arguments of the \code{pdata}, \code{ndata},
\code{nout} and \code{pout} variables structures that will be passed to
\code{particleParticleInteraction}.

The most valuable upside to this single-parameter approach to function templates is its simplicity,
without any loss of expressive power.
This comes at the cost of reduced type checking.
It would indeed be possible to pass \emph{anything} to the function templates mention
before, even something that is not a specialization of the expected template
(e.g. something that is not a \code{forces\_params} specialization for \code{forces}
in Listing~\ref{l:constexpr}).

This is however for the most part only a theoretical issue,
as in most cases the instantiation of the template or of any of the auxiliary functions
will trigger a compilation error when trying to access a non-existent member
for anything but a specialization of the correct structure.
Moreover, the possibility to plug in almost arbitrary data structures for these
generically-specified arguments structures actually gives more flexibility, since
it allows the same function templates to be used with potentially unrelated structures
that still present the same (or similar) sets of member variables,
thereby improving the genericity of the code by supporting a form of \emph{duck typing}.
This is actually a benefit, especially for the smaller functions down the call chain
that may get passed variables structures from different computational kernels.

For example, some density diffusion contributions may be computed during the forces computation
(thus receiving the variables structures from the \code{forces} kernel)
or as a separate post-integration step
(thus receiving the variables structures from a dedicated \code{densityDiffusion} kernel),
depending on the scheme being employed. Duck typing allows the same device function
to compute the contribution in both cases, provided the members of the variables structures
have the same name and meaning.

In fact, the only \emph{practical} downside we have found to this approach has been,
as developers, the sometimes overwhelming complexity of the error messages when
the instantiation of a specialized version of a structure or function template fails.
Template errors in C++ are notoriously long-winded and hard to decipher, to the point
of being commonly classified as \emph{horrific}. This issue is compounded, in CUDA,
by a paradoxical lack of \emph{useful} information from the \code{nvcc} compiler
in some instances
(particularly in combination with the SFINAE tricks we will discuss momentarily),
a fact that was actually one of the driving forces for exploring alternative compilers,
with the surprising results discussed in section~\ref{sec:clang}.

\subsection{SFINAE for function template specialization}
\label{sec:sfinae}

The remaining C++ programming feature employed in our effort to improve the genericity of the code
pertains the specialization of function templates with several template parameters,
but for which the specialization depends on a single parameter.

Assume for example we have a function template \code{f} depending
on the boundary type and turbulence model,
for which we would like to write a specialization for the case
of the $k$-$\epsilon$ turbulence model, regardless of the boundayr type.
Since C++ does not support partial function template
specialization, something like
\begin{lstdisplay}
template<BoundaryType, TurbulenceModel> void f(/* omissis */);
template<BoundaryType boundarytype> void f<boundarytype, KEPSILON>
	( /* omissis */ )
\end{lstdisplay}
would be invalid syntax.


The idiomatic way to achieve the intended effect
(that we have extensively adopted in GPUSPH version~5)
relies on the C++ feature known as
Specialization Failure Is Not An Error (SFINAE).
According to the language specification, when trying to resolve a function name,
any function template for which template parameter deduction would result in an error is simply ignored.
This allows a form of partial specialization to be achieved by \emph{overloading}
the function template, but allowing one, and only one overload to resolve correctly
based on the intended specialization.

The fundamental building block for this is a structure template that declares
the actual return type of the function template in the positive case,
but does not declare anything as return type (and thus results in an error, and the discard
of the overload) otherwise.

In C++11, this type is provided by the standard library \code{std::enable\_if},
a structure template such that \code{std::enable\_if<true, T>::type}
will resolve to the type name \code{T}, whereas \code{std::enable\_if<false, T>::type}
will result in an error (the type name \code{T} can actually be omitted,
in which case \code{void} is assumed by default).
A similar structure can be defined by hand also in earlier revisions of the C++ language:
\begin{lstdisplay}
template<bool B, typename T=void>
struct enable_if {}; /* no `type` member, to cause an error */

template<typename T>
struct enable_if<true, T> { typedef T type; };
\end{lstdisplay}
but C++11 brings with itself
not only the definition out of the box,
but also the possibility to define a type alias template to simplify syntax:
\begin{lstdisplay}
template<bool B, typename T=void>
using enable_if_t = typename enable_if<B, T>::type;
\end{lstdisplay}
so that \code{enable\_if\_t<bool, T>} can be used as a shorter form for \code{enable\_if<bool, T>::type}.
This type alias template is predefined since C++14.

Partial specialization can then be achieved by defining multiple overloads of the function templates,
with a return type in the form \code{enable\_if\_t<condition, actual\_return\_type>},
and using mutually exclusive conditions for each of the partial spcializations.
For example, assuming that \code{void} is the return type of our \code{f} function template above,
we can use:
\begin{lstdisplay}
template<BoundaryType boundarytype, TurbulenceModel turbmodel>
enable_if_t<turbmodel == KEPSVISC> f(...)
{ /* code for the KEPSVISC case */ }

template<BoundaryType boundarytype, TurbulenceModel turbmodel>
enable_if_t<turbmodel != KEPSVISC> f(...)
{ /* code for the non-KEPSVISC cases */ }
\end{lstdisplay}
to fulfill our need to specialize \code{f} for the $k$-$\epsilon$ turbulence model.

This feature can be combined very efficiently with the template argument reduction discussed
in section~\ref{sec:fp}:
\begin{lstdisplay}
template<typename AV>
enable_if_t<AV::turbmodel == KEPSVISC> f(AV& args)
{ /* code for the KEPSVISC case */ }

template<typename AV>
enable_if_t<AV::turbmodel != KEPSVISC> f(AV& args)
{ /* code for the other cases */ }
\end{lstdisplay}
with the added benefit of causing a compiler error if an incompatible argument type
(specifically, one without a \code{turbmodel} static constant member)
is passed to \code{f}, since the compiler will discard both specializations
(with the SFINAE-ignored failure being the inability to access \code{turbmodel})
and complain about being unable to resolve the function name.

\section{Putting it all together: split neighbors}

The adoption of the meta-programming techniques discussed so far has been the key
to one of the most significant performance boost from version~4 to version~5 of GPUSPH:
the split neighbors list processing, which has brought a typical performance improvement
between 15\% and 30\%, depending on the combination of framework options
and hardware capabilities~\cite{bilotta_spheric_2019,bilotta_design_2018}.

Analysis of the performance of the \code{forces} computational kernel in version~4
revealed that, despite the high density of computational operations in the kernel,
its runtime was still largely memory-bound. The main cause for this was tracked down
to the large number of variables that had to be allocated during the processing
of the particle-particle interactions, which resulted in them overflowing the
register banks of the GPU multiprocessors, resulting in the usage of
the much slower VRAM as temporary storage (termed \emph{local memory} in CUDA).

This large register pressure was ascribed to the monolithic nature of the kernel
and the disparity of behavior in the interaction between particles of different types:
indeed, for most boundary models fluid-fluid particle interactions are different from
fluid-boundary particle interactions, but since particles are stored all together,
and the neighbors list stored all neighbors together (without distinction of type),
during execution of the monolithic kernel each work-item would
load the data for the particle being processed,
and then traverse the entire neighbors list, deciding \emph{at runtime} how
to interact with each specific neighbor, based on the neighbor type.

This led to a growth in the register usage, since the total number of variables
that needed to be allocated was no less than the \emph{union} of the variables needed
for each of the particle-particle interaction types. This approach also led
to additional performance loss due to the runtime decision about the kind of interaction,
and the possible divergence of execution code-paths
due to disparity of interaction between pairs of particles being processed concurrently,
a well-known performance issue on GPUs.

\begin{lstfloat}%
[caption={Full template signature for the \code{forces} computational kernel as of version~5 of~GPUSPH,
showing also the use of optional template arguments as a shortcut to expose the static constant members
of the arguments structure template \code{forces\_params} (in order to make them accessible
within the function body without the \code{FP::} prefix),
and the addition of the central (\code{cptype}) and neighbor (\code{nptype}) particle type
as template parameters (in \code{forces\_params} and inherited in \code{forces})
for the split-neighbors implementation.},
 label={l:realforces}]
template<typename FP, /* forces_params<kerneltype, sph_formulation, etc> */
	KernelType kerneltype = FP::kerneltype,
	SPHFormulation sph_formulation = FP::sph_formulation,
	DensityDiffusionType densitydiffusiontype = FP::densitydiffusiontype,
	BoundaryType boundarytype = FP::boundarytype,
	typename ViscSpec = typename FP::ViscSpec,
	flag_t simflags = FP::simflags,
	ParticleType cptype = FP::cptype,
	ParticleType nptype = FP::nptype>
void forcesDevice(FP params)
\end{lstfloat}

The solution we adopted has been to split the \code{forces} kernel into a \emph{separate}
specialization for each of the particle-particle interaction pairs: one for fluid to fluid,
one for fluid to boundary, one for boundary to fluid, etc.
Hence, the \code{forces\_params} parameters structure template and all the variables structure templates
for the \code{forces} computational kernel depend not only on the framework parameters,
but also on the central and neighbor particle type, as illustrated in Listing~\ref{l:realforces}
that shows the declaration of the function as of versio~5 of~GPUSPH.

The split has been particularly
meaningful for the semi-analytical boundary conditions~\cite{ferrand_2012,mayrhofer_2015},
that have three different particle types, larger disparities in the treatment of different pairs,
and complex rules to decide which pairs should be computed: even just moving most of
the decision logic outside of the device-side computational kernel to the host-side
has given a measurable performance gain of a few percents.

Reducing the complexity of the logic inside each instance of the computational kernel
(now one per pairwise combination of particle types)
gives more optimization opportunities to the compiler, especially when most of the conditions
end up depending only on template parameters, whose value is known at compile time.
On the other hand, by itself this strategy is insufficient,
since leaving the kernel body and associated data structures unchanged leaves two main reasons of inefficiency.

The first comes from the need to issue each specialization of the kernel on the entire set of
particles, with each specialization then skipping the particles whose type is different from
the requested one. For example, the fluid-fluid and fluid-boundary versions would skip non-fluid particles.
This can be solved by creating a separate particle system for each particle type,
and issuing each version of the kernel only on the particle system of the correct type.
However, since fluid particles are generally an order of magnitude more than the other
particle types, the impact of the loss is relatively low, while the implementation cost
of the solution is quite high, so that optimizing this aspect has been considered ``second-order'',
and therefore postponed.

The second inefficiency comes from the traversal of the neighbors list: when all neighbors
are stored together, regardless of type, it will be necessary to randomly skip elements
from the neighbors list during its traversal. This can be quite costly, as it is again a likely source of
divergence at the hardware level.
We have solved the issue by redesigning the way neighbors are stored in the list, in such a way that
all neighbors of the same type are stored consecutively, while preserving
GPU-optimal access patterns and without increasing the storage requirements.

\subsection{Efficient split neighbors list}
\label{sec:split-neibs}

Memory access is a significant bottleneck on most modern computational hardware, be it CPUs or GPUs.
This is the main reason for the growing, multi-layer caches of CPUs, and the introduction of
L1 and L2 caches even on GPUs. Optimal memory access patterns can significantly speed up an implementation,
but the optimality of the patterns depends not only on the nature of the algorithm,
but also on the characteristics of the hardware.

On a mostly sequential processor like a CPU, the best cache usage is obtained by
placing the data needed by a single thread in adjacent memory locations.
By contrast, on stream processing hardware like GPUs, optimal access patterns require
that adjacent memory locations refer to data needed concurrently by different work-items.

Let us consider the case of the neighbors list, and let us denote by $n_i(j)$
the index of the $i$-th neighbor of the $j$-th particle.
When traversing the neighbors list, a thread or work-item processing particle $j$ will need to read
the index of the first neighbor ($n_0(j)$), then
the index of the second neighbor ($n_1(j)$), and so on.

For sequential hardware (such as CPUs), it is then optimal to lay out the neighbors indices in memory grouping them
by particle:
\[
n_0(0), n_1(0), \dots, n_k(0), n_0(1), n_1(1), \dots, n_k(1), \dots
\]
where $k+1$ is the maximum number of neighbors per particle.
This will ensure that whenever a neighbor index is loaded from memory,
the following neighbors indices will be cached too.

(Note: for efficiency reason, we use a fixed-size neighbors list, so that the neighbors of each
particle can be found by simple computations, without additional memory accesses. A special marker
is added to the list of neighbors of each particle if it's not full, to indicate the actual end.)

By contrast, on GPUs we want to lay out the list in memory in such a way that when work-items
$j, j+1, \dots$ load the index of their respective first neighbor from memory, they will find it
in adjacent memory locations. In this case it is therefore optimal to lay out the neighbors indices in memory
grouping them by ordinal:
\[
n_0(0), n_0(1), \dots, n_0(p), n_1(0), n_1(1), \dots, n_1(p), \dots
\]
where $p$ is the number of particles.

\begin{figure*}
\centerline{\includegraphics[width=\columnwidth]{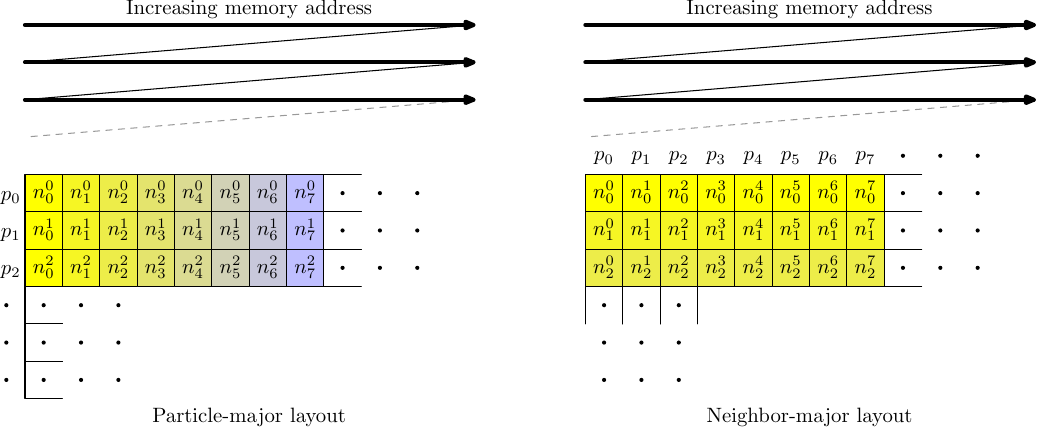}}
\caption{Different layouts for the neighbors list, comparing the CPU-preferrable ``by particle''
to the GPU-preferrable ``by neighbor'' order.}
\label{fig:neibs-layout}
\end{figure*}

We call this layout \emph{interleaved} or \emph{neighbor-major},
in contrast to the \emph{sequential} or \emph{particle-major} layout used for CPUs,
since from the perspective of each particle, the next neighbor is not found with an offset $+1$, but
with an offset $+p$, with the memory in-between dedicated to the same neighbor ordinal for the other particles
(Figure~\ref{fig:neibs-layout}).

This works very well when the entire neighbors list must be traversed each time, but is sub-optimal
when we want to only traverse a subset of the list
(for example, as we do in GPUSPH, considering only the neighbors of a given type).

To illustrate the approach we adopted, assume at first that we have only two particle types (fluid and boundary),
and that when there are more neighbors of one type, there will be fewer of the other type. This is consistent
with the fact that a particle far from the boundary will have a full neighborhood of fluid particles,
but as it gets closer to the boundary the number of fluid neighbors will decrease, while the number of boundary
neighbors will increase at a similar rate (assuming a more-or-less uniform particle distribution).

The solution to the split neighbors list in this case is to collect all neighbors of one type at the beginning
of the list (from the first location up), and all the neighbors of the other type at the \emph{end} of the list
(from the last location down). If we denote by $f_i$ the $i$-th fluid neighbor and $b_i$ the $i$-th boundary neighbor,
from the perspective of a single particle, the neighbors would then be stored as:
\[
f_0, f_1, f_2, \dots, f_{k_f}, \odot, \dots, \odot, b_{k_b}, \dots, b_2, b_1, b_0
\]
where $k_f+1$ is the number of fluid neighbors, $k_b+1$ the number of boundary neigbors, and $\odot$ denotes
the end-of-list marker.
Of course this single-particle perspective can then be implemented system-wide using either the sequential layout,
or with the interleaved layout.

\begin{figure}
\centerline{\includegraphics[width=.5\textwidth]{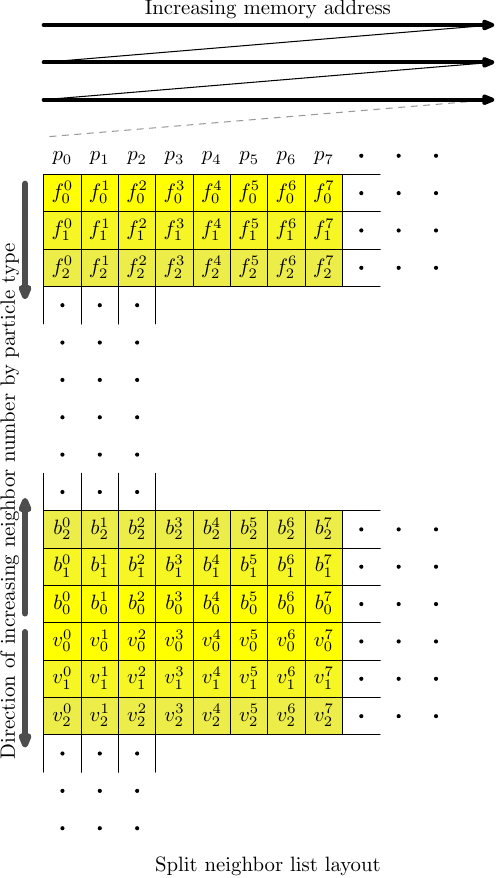}}
\caption{Split-neighbors-list layout in the neighbor-major order case.}
\label{fig:splitneibs-layout}
\end{figure}

The situation is more complex when the number of particle types grows. In the case of a third type,
for example (as is the case for the semi-analytical boundary conditions with their vertex particles),
the neighbors list will need to be split into two fixed-size chunks, one to store the first pair of types,
and the other to store the remaining type (Figure~\ref{fig:splitneibs-layout});
indicating as before with $f_i, b_i, v_i$ respectively the $i$-th fluid, boundary and vertex neighbor,
and with $k_v$ the number of vertex neighbors, the list as seen by each particle would be coded as:
\[
f_0, f_1, f_2, \dots, f_{k_f}, \odot, \dots, \odot, b_{k_b}, \dots, b_2, b_1, b_0,
v_0, v_1, v_2, \dots. v_{k_v} \odot
\]
In this case, it's necessary to know the size of the chunk. In GPUSPH, we store the (local) index
of the first boundary neighbors index, knowing that the vertex neighbors will start at the next location.

\section{Textures versus linear memory}
\label{sec:tex}

One of the key differences between GPUs and CPUs is the possibility to access global memory
not only through the standard linear addressing mode, but also through the texture mapping units (TMUs),
dedicated hardware that supports a number of features such as multi-dimensional addressing,
hardware interpolation and type conversion (e.g. from integer type to normalized floating-point values).

On older GPU architectures, using textures was also the only way to have some kind of caching,
due to the TMU approach of loading a small neighborhood of the pixel values when fetching the data for a single pixel.

Judicious use of textures and linear memory addressing has thus been a sure-fire way in GPUSPH to maximize memory throughput,
even after the introduction of the L1 and L2 caches in more modern GPU architectures, at least until the introduction of
a unified texture and L1 cache in the Pascal architecture.

Since linear and texture memory addressing requires different function calls,
and the distribution of the data arrays between the two in GPUSPH follows a complex logic based on
a combination of number of arrays and hardware characteristics,
the code to fetch the particle data during a kernel can become rather complex.

The use of the argument structures (section~\ref{sec:cond-struct}) allows the simplification of this logic
by adding appropriate members to the structure:
for example, the substructure holding the particle position (resp. velocity) array
can provide a \code{fetchPos(index)} (resp. \code{fetchVel(index)}) member function
that maps to a linear memory access or a texture access depending on the caching preference.
When all the substructures are combined in a single argument structures, the kernel can then use
\code{params.fetchPos(index)},\code{params.fetchVel(index)} etc
without having to worry about the details of how the data is stored.

This is even more convenient for data types (such as symmetric tensors) that do not map
to a native vector data type and as discussed in Section~\ref{sec:sph-impl} may be split
over multiple arrays in device memory: a \code{params.fetchTau(index)} function
can read the three \code{float2} values from the split arrays in \code{params}
and assemble the full tensor to be returned to the caller.

Moreover, this also allows changing the storage logic without having to change any of the kernels accessing the data,
a feature that has significantly helped us in supporting alternative compilers
whose support for texture memory is still insufficient, as discussed in section~\ref{sec:clang} below.

\section{Downsides}
\label{sec:down}

The strategies discussed in this paper present significant advantages both in terms of code clarity
and in terms of performance. However, these benefits have some costs, that will be described in this section.

The first price to be paid is the reduced approachability to the code for less experienced developers.
Any complex code base suffers from similar issues, but in the specific case of GPUSPH the difficulty lies
in the limited exposure most developers have had to these more advanced programming strategies,
which are thus less familiar than more traditional approaches.

The only reliable solution to this problem is extensive documentation. An effort has been made to
improve the in-code documentation to this end, that is being complemented by a more comprehensive,
and easier to peruse, reference manual to assist researchers interested in extending GPUSPH.

The second price, paid also by more experienced developers and even those familiar with the GPUSPH code already,
is the difficulty in associating the generic template parameter names with the corresponding data structures.
This is important when extending the underlying structures to include new (optional) members.

Since we want to preserve the high flexibility of the function templates, the only solution in this case too
is to improve the in-code documentation, although we are also exploring alternative approaches that would
allow an automatic tracking of ``which structures get passed to which functions''.
Since this is not possible out-of-the-box with the standard CUDA compiler \code{nvcc},
this will require to look into more flexible compiling solutions.

The final price to be paid is the already mentioned increase in verbosity of error messages from the compiler
during development, due to the heavy usage of function and structure templates.
This is compounded by a surprising lack of information from the standard CUDA compiler in case of failure
of selection of a function template overload.
Specifically, \code{nvcc} will inform the developer about the failure to find the correct instance,
and it will mention which overloads were considered, but it will provide no information about \emph{why}
each of the considered overloads was excluded from the selection,
which makes it much harder to fix any developer mistake e.g. in the conditionals used for \code{enable\_if}
(a common cause of such errors).

This last issue in particular was the main reason why we started looking into alternative compilers,
and in particular Clang~\cite{clang},
that has been working on built-in support for CUDA for several years~\cite{gpucc,clang_cuda}.

\section{Clang: the better compiler?}\label{sec:clang}

Clang~\cite{clang} is a compiler front-end to LLVM~\cite{llvm_2004} for the C family of languages
whose development was started by Apple to complement its investment in LLVM and to detach themselves from GCC,
due to the complexity and stricter licensing of the latter~\cite{clang_new}.

Since its publication as open source software in 2007, Clang and LLVM have gained a lot of momentum,
not only from commercial vendors (Apple, Microsoft, Google, ARM, Intel, AMD),
but also among researchers and free-software developers~\cite{pocl,mesa}
due to a combination of more liberal licensing, faster compilation time,
easier integration with other software and more developer-friendly error messages.

(Interestingly, around the release of CUDA 4, even NVIDIA transitioned from their old
Open64-based compiler to a new LLVM-based compiler. However, given the results that will be discussed below,
it's clear that their fork has not been able to keep up with the progress of the
upstream Clang/LLVM combination.)

The boost in development has extended both the hardware support for binary code generation in LLVM
(covering not only traditional CPUs, but also several GPUs from a variety of vendors)
and the languages supported by the Clang front end
(C, C++, OpenCL, RenderScript, CUDA, SYCL).

The recent improvements for Clang's support for CUDA~\cite{gpucc,clang_cuda}
and the growing complexity of the GPUSPH code,
with the resulting difficulty in understanding the cause of some compiler errors
during development,
have been the driving forces behind our exploration of the use of Clang to compile GPUSPH,
as attested by the first commit\footnote
{GPUSPH commit hash a3d5913b4d0bc55bd8cbc7c0847fbdc5234a63dc, authored on 2019-03-06}
in the GPUSPH history that introduced Clang support:

\begin{quote}
Preliminary support for using Clang's CUDA support

This is mostly used for debugging, since Clang provides much better
error messages and code analysis.
\end{quote}

The biggest obstacle to the use of Clang in GPUSPH was the extensive use of textures
to improve cache efficiency (section~\ref{sec:tex}), since Clang's support for textures in CUDA
was at the time (and still is, in some circumstances) unreliable, sometimes even failing in the binary code generation phase.
This was solved by making all texture access optional, and adding a preprocessor define
to disable textures completely: this is enabled by default when using Clang, but can also be enabled
when building with \code{nvcc}, allowing a more fair comparison between the compilers.

(As it turns out, on the more recent GPUs there is no measurable difference in performance
between non-interpolated texture and linear memory usage, so the test-cases affected by the switch
on such hardware are only those using a digital elevation model.)

When the first Clang-compiled test case was run,
a simple three-dimensional dam break,
it was therefore quite surprising to see a nearly $1.5\times$ performance boost over the \code{nvcc}-compiled code%
.
The performance gain was confirmed across multiple hardware generations and compiler versions,
with tests spanning Clang version~9 to~12 and \code{nvcc} versions 10.1 to 11.4.

In what follows we will describe the analysis that led to the discovery of the source of this discrepancy,
the changes introduced to allow a similar performance gain across compilers, and the side-effect this had
on multi-GPU support. Detailed benchmarks on the net effect of these improvements will be presented
in section~\ref{sec:benchmarks}.

\subsection{Analysis of the performance difference}

\begin{figure}
\includegraphics[width=\textwidth]{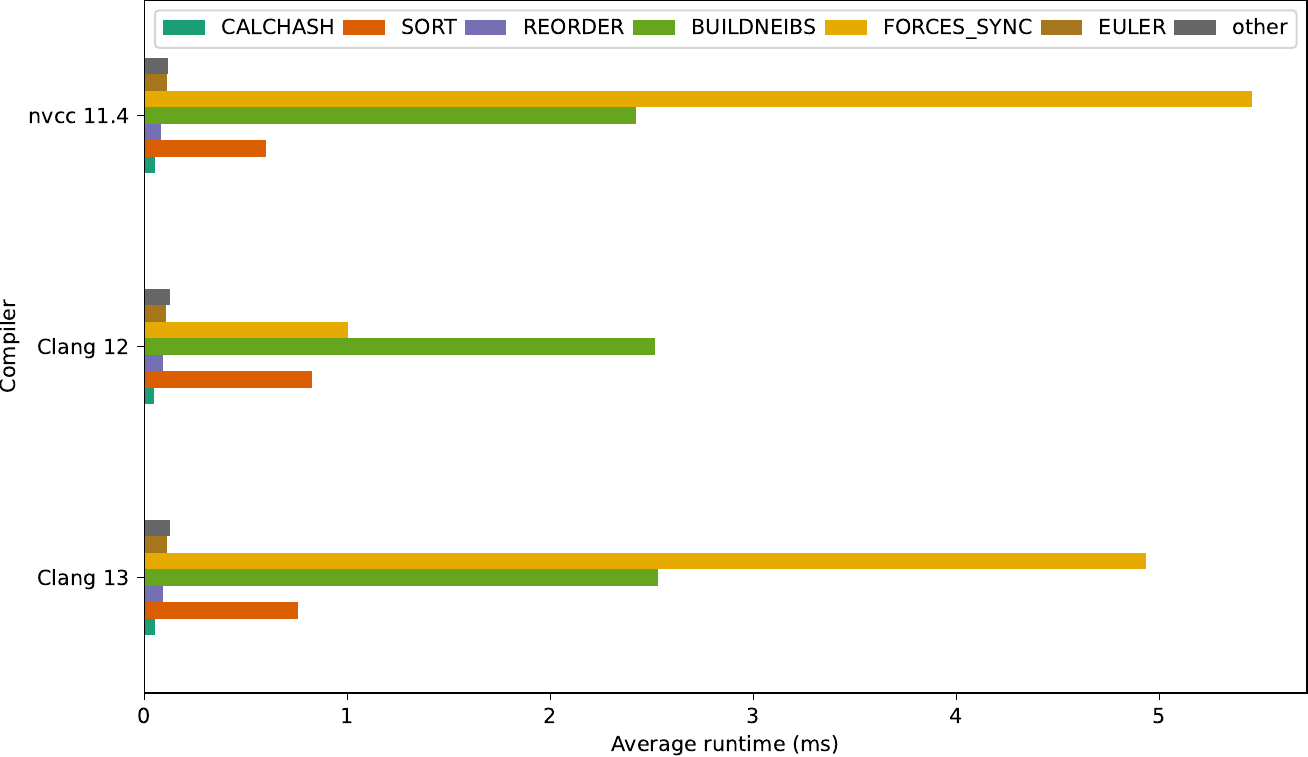}
\caption{Average runtime (in ms) per command invocation of the largest contributors, by compiler, with the initial code.}
\label{fig:avg-before}
\end{figure}

\begin{figure}
\includegraphics[width=\textwidth]{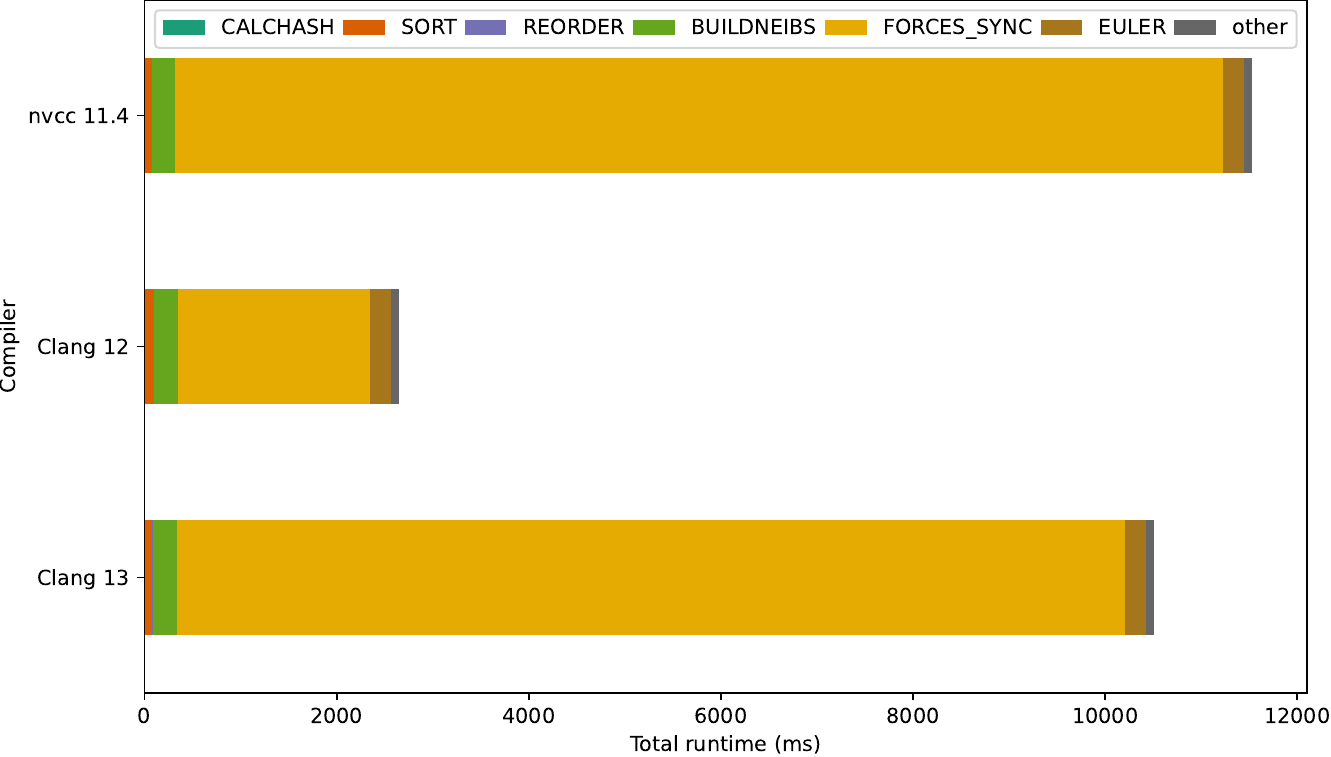}
\caption{Total cumulative runtime (in ms) for each command during the first 1000 simulation steps
of the performance analysis test-case, by compiler, with the initial code.}
\label{fig:tot-before}
\end{figure}

\begin{figure}
\includegraphics[width=\textwidth]{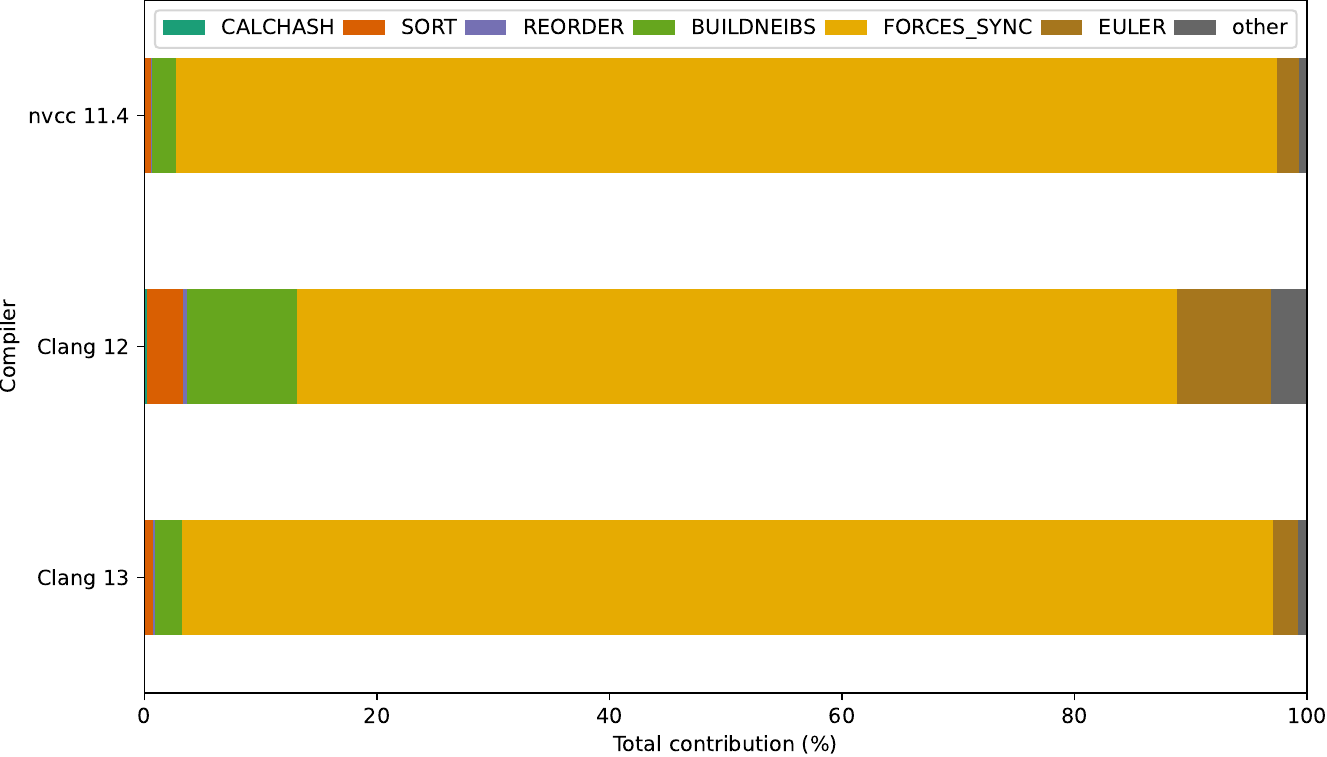}
\caption{Percent contribution by each command to the total runtime during the first 1000 simulation steps
of the performance analysis test-case, by compiler, with the initial code.}
\label{fig:totpct-before}
\end{figure}

The test case used for the comparison is a three-dimensional dam-break with in a $1.6\text{m}\times 0.67\text{m}\times0.6\text{m}$ domain.
The initial water volume occupies one side of the box, up to a height and depth of $H=0.4$m.
Framework options include Lennard--Jones boundary conditions for the solid walls, 
artificial viscosity,
and the Molteni \& Colagrossi~\cite{molteni_colagrossi_2009} density diffusion model.

The simulations,
consisting of $53,248$ fluid and $30,388$ boundary (for a total of $83,648$) particles at a resolution of 32 particles in the $H$ length (\emph{ppH}),
were only run for $1,000$ iterations, to gather basic information about total runtime and per-command contributions
(Figures~\ref{fig:avg-before}--\ref{fig:totpct-before}).
Data was not saved, since we were only interested in the computational performance.
The results illustrated in the plots Figures~\ref{fig:avg-before}--\ref{fig:totpct-before} refers to
execution on an NVIDIA GeForce GTX 1650 Max-Q.

The first comparison was done between \code{nvcc} 11.4 and Clang 12.
It was observed (Figures~\ref{fig:avg-before} and~\ref{fig:tot-before}),
that many commands had in fact a marginal performance \emph{regression} with Clang,
with the most significant exception being the \code{FORCES\_SYNC} command that runs the forces computation kernels,
that take up the lion's share of the simulation (Figure~\ref{fig:totpct-before}),
thanks also to the fact that \code{FORCES\_SYNC} and \code{EULER} are run twice per time-step,
while the ancillary commands for neighbors list construction
(\code{CALCHASH}, \code{SORT}, \code{REORDER}, \code{BUILDNEIBS}),
are only run once every 10 time-steps.

A second surprising result came with the release of Clang~13,
testing on which resulted in performances within less than $20\%$ of those obtained
when comping with \code{nvcc}
(sometimes in excess, sometimes in defect, depending on test case and GPU architecture).
Compared to the consistently improved performance of Clang~12
the results from Clang~13 were considered a regression in the compiler,
and reported as an issue to the Clang developers~\cite{clang13_regression}.

A more thorough analysis of the forces kernels revealed that the key to the performance
differences was in the usage of the stack.
On~GPU, use of the stack is particularly nefarious, as it involves the use
of appropriately reserved global memory, which is no less than two orders of magnitude slower
than registers, and can introduce significant latency in hot-path code.
Two elements were in fact surprising about the stack usage reported for the forces (and many other) kernels:
the first was that, since all device functions are marked with the \code{always\_inline} attribute in GPUSPH,
there \emph{should} have been no stack usage at all;
and the second was the question why Clang up~to version~12 managed to avoid it,
whereas the other compilers (all \code{nvcc} version and Clang~13 and higher) required it.

While the latter remains a mystery to date, resolved in Clang~14
by introducing an additional optimization pass in the compilation of CUDA code,
the first question was answered by discovering the culprit
in the virtual inheritance involved in the definition of the neighbors list iterators.
Avoiding this was key to providing a significant performance boost across compilers.

\subsection{The return of the neighbors list: multi-type iterators}

The split neighbors list described in section~\ref{sec:split-neibs} makes it very efficient
to traverse the list of a single neighbor type.
An iterator simply needs to load the beginning of the chunk and the traversal direction
(based on the neighbor type), and then load each neighbor index in sequence until the
end-of-list marker is encountered.

When the same interaction needs to be computed with neighbors of different types
(as is the case frequently with dynamic boundary conditions),
the code necessary to traverse the neighbors list is made more complex by the need to
change the offset and direction of traversal when one type is finished and the other begins,
but aside from that the behavior remains essentially the same.

To abstract all this from the developer, we implemented \code{neibs\_iterator} class templates
that take care of all the internals, and present a simple interface with methods to
retrieve the current neighbor's index, its relative position to the central particle,
and finally a method to fetch the next neighbor and inform the caller when the
neighbors list is completed.

The core of all the iterators is the same: a set of variables independent of the particle type,
and the internal methods to fetch and decode the neighbor information from the neighbors list.
These are abstracted in a dedicated \code{neibs\_iterator\_core} class.

All the single-type \code{neibs\_iterator} class templates derive from the core class,
and simply implement on top of it the necessary detail for the specific neighbor type,
such as the computation of the chunk start offset and the traversal direction.

\begin{figure}
\includegraphics[width=\textwidth]{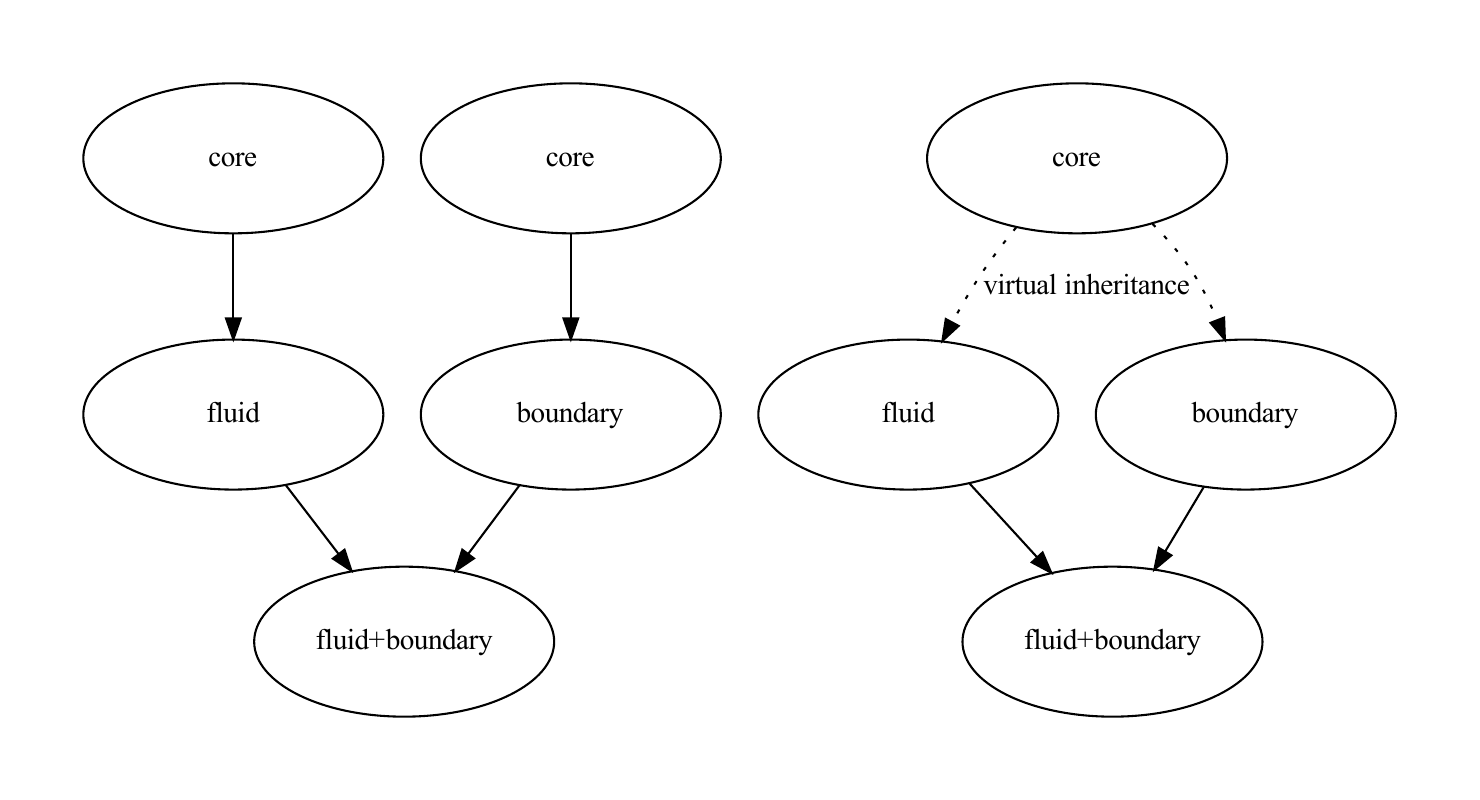}
\caption{Diamond inheritance of base classes with and without virtual inheritance through
intermediate classes. In the standard case (left), two copies of the core class
are inherited by the multi-type iterator, making any reference to \code{neibs\_iterator\_core}
ambiguous. With virtual inheritance (right), a single copy of the core class is inherited,
making it possible to access \code{neibs\_iterator\_core} unambigious
in the multi-type iterator.}
\label{fig:virtual-diamond}
\end{figure}

The straightforward way to implement a multi-type iterator is to create a dedicated class
that derives from the corresponding single-type iterators, and switches from one to the other
as the end-of-list marker is reached. This however leads to the infamous \emph{diamond inheritance}
problem: since all the single-type iterators depend on the core class, and we want a single copy
of the core class as (grand)parent of the multi-type iterator,
the single-type iterators have to declare the core class as a \emph{virtual} parent
(Figure~\ref{fig:virtual-diamond}).

This virtual inheritance, however, is in our case responsible for the inefficiency
experienced with \code{nvcc} and Clang~13, as the compiler fails to de-virtualize the structure.
Even worse, the negative impact of the virtual inheritance affects single-type iterators too,
even though for them there would be no need for virtualization in the first place:
indeed, single-type iterators inherit virtually from the core class only to support multi-type iterators.
This is actually in conflict with one of the principles behind the GPUSPH design,
that the implementation of a feature should not negatively affect other features
(in this case, the possibility to iterate over multiple types should not affect
the performance of iterating over a single type).

\begin{figure*}
\includegraphics[width=\textwidth]{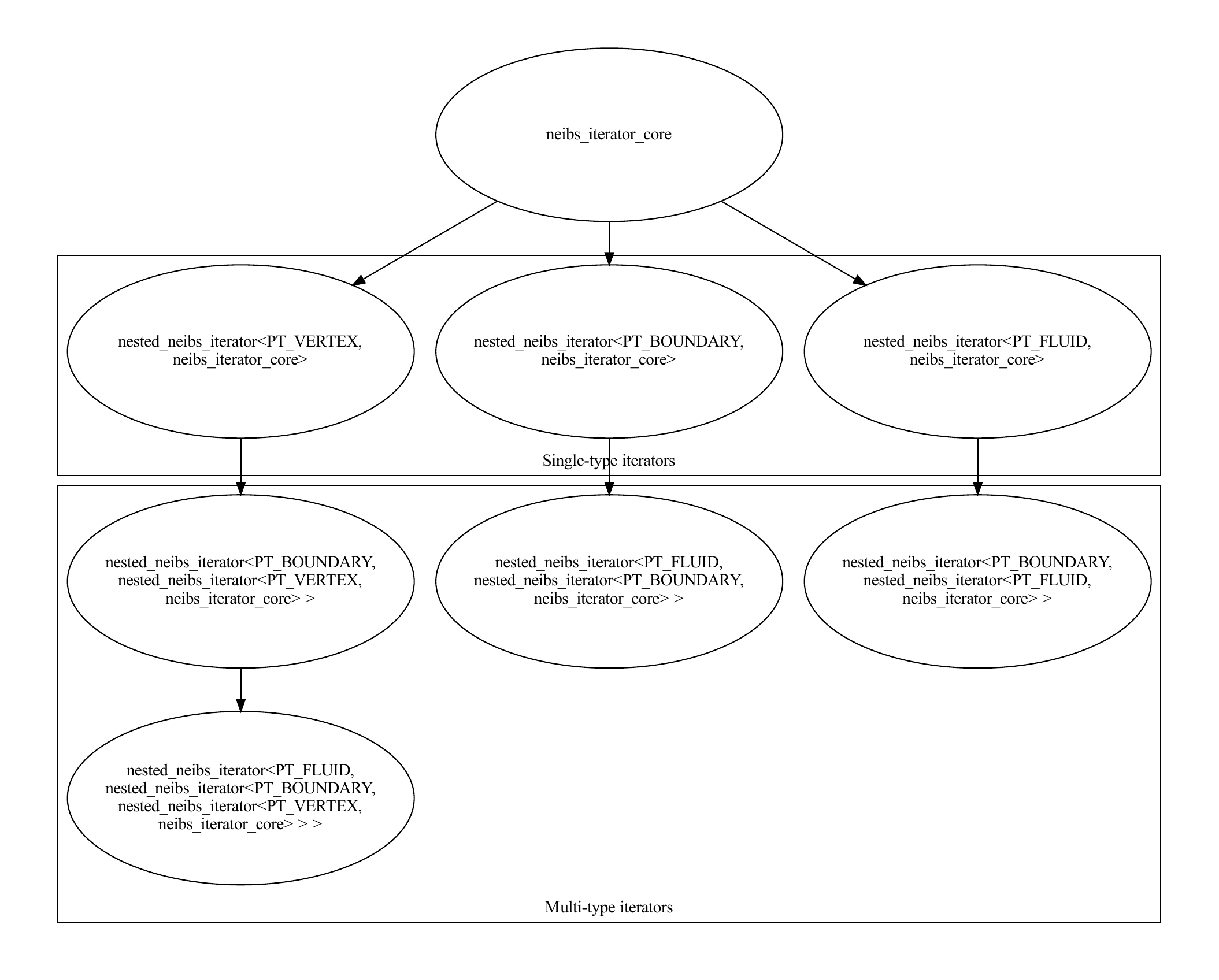}
\caption{Chain inheritance through nesting class templates.
Single-type iterators derive directly from the core class,
multi-type iterators derive from the other multi- or single-type iterators.
For multi-type iterators, the order of the nesting determines the order in which neighbor types are processed.
On the middle, for example, fluid types are processed before boundary types,
whereas on the right boundary types are processed first.}
\label{fig:chain-inheritance}
\end{figure*}

The approach we adopted in GPUSPH to avoid virtual inheritance and its negative effects
on GPU performance has been to turn multiple inheritance into \emph{chain} inheritance,
leveraging the fact that we iterate over neighbor types in a predefined order
(Figure~\ref{fig:chain-inheritance}).

Instead of distinguishing between single- and multi-type iterators, we define a single
class template for all iterators, with two template parameters:
the particle type of the `current' iterator, and the class of the `next' iterator,
with the template defining a class that derives from the `next' iterator:
this base class is then used to delegate the retrieval of the next
neighbor when the current type is exhausted (Listings~\ref{l:nested-iterator}).

\begin{lstfloat}%
[caption={Nesting class template to implement multi-type neighbors list iterators.},
 label={l:nested-iterator}]
template<ParticleType ptype, typename NextIterator>
class nested_neibs_iterator : public NextIterator
{
	using core = neibs_iterator_core;

	bool next() {
		if (core::current_type == ptype) {
			// fetch the next neighbor of this type
			bool has_next = fetch_next_neighbor();
			if (has_next) return true;
			// no more neighbors: switch to next type
			NextIterator::reset();
		}
		// this type has finished, delegate to next iterator
		return NextIterator::next();
  	}

	/* rest of the class omitted for brevity */
};
\end{lstfloat}

\begin{figure}
\includegraphics[width=\textwidth]{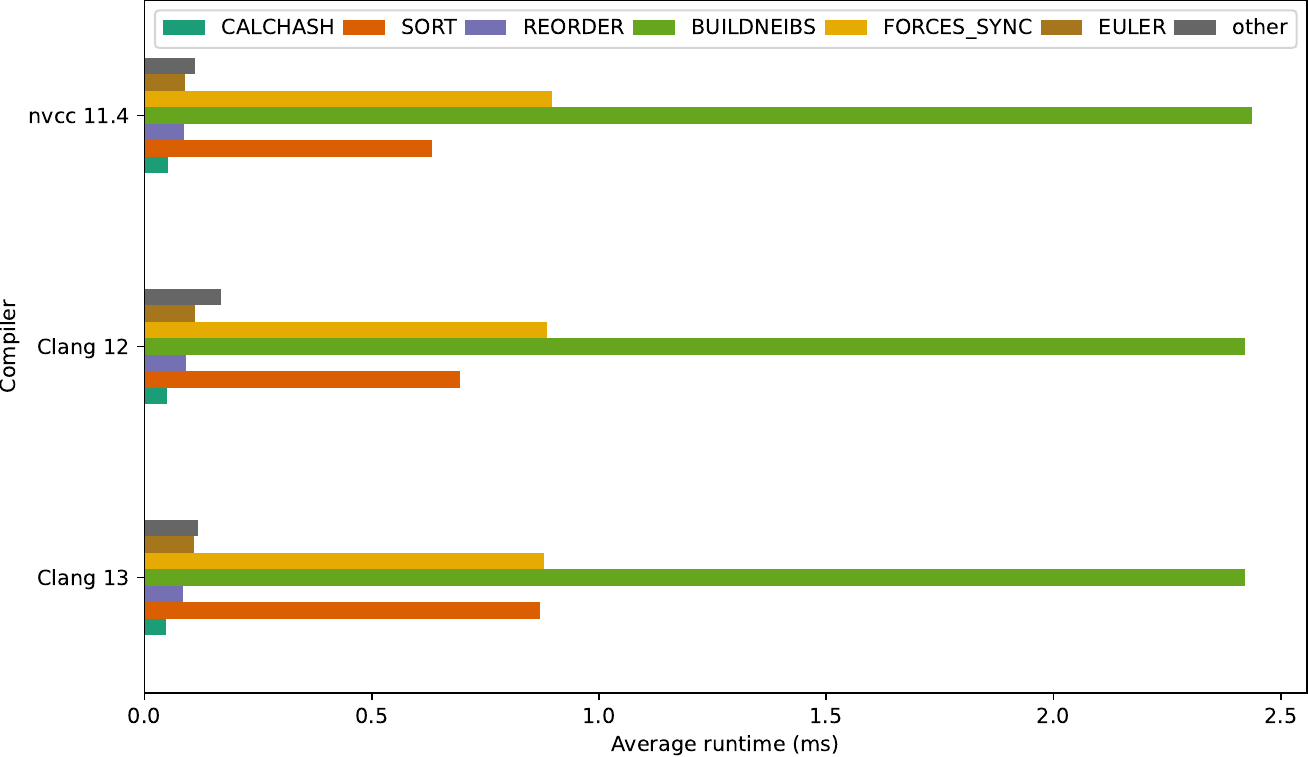}
\caption{Average runtime (in ms) per command invocation of the largest contributors, by compiler, with the devirtualized iterators.}
\label{fig:avg-after}
\end{figure}

\begin{figure}
\includegraphics[width=\textwidth]{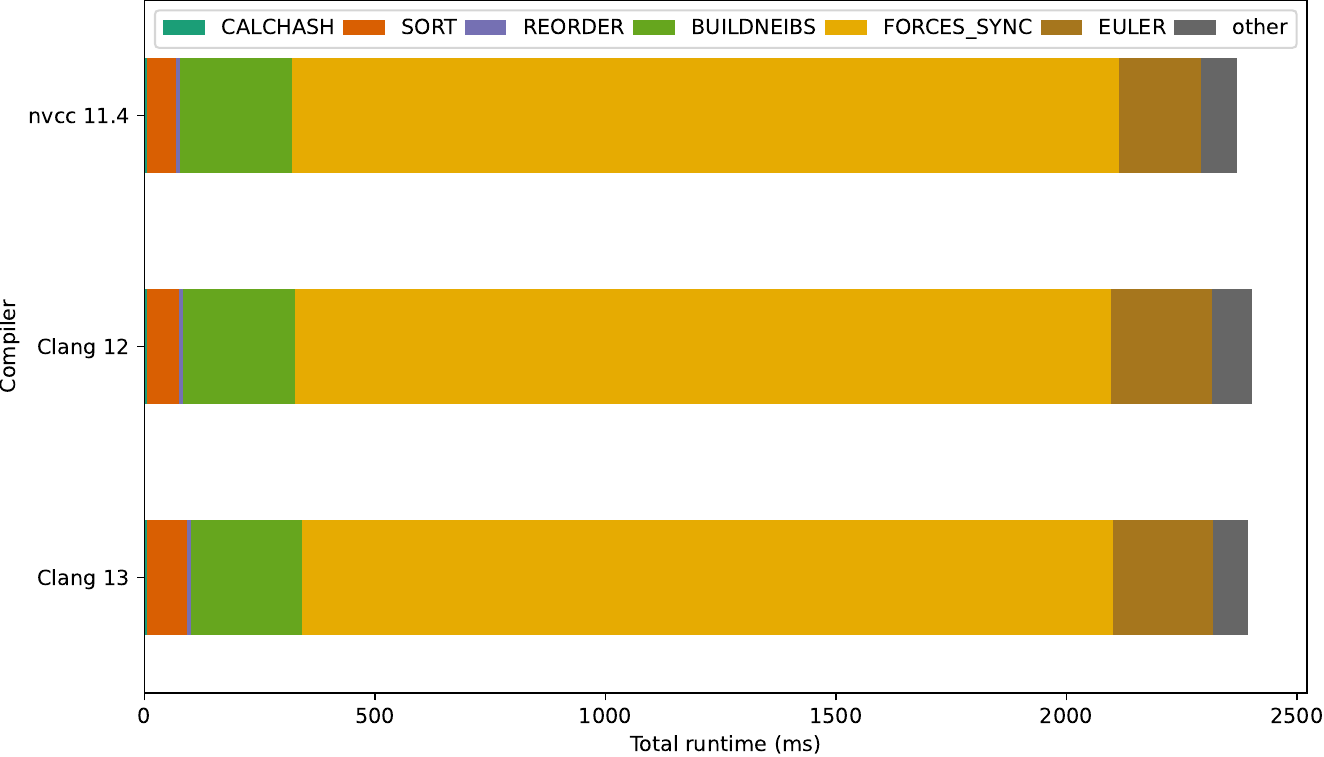}
\caption{Total cumulative runtime (in ms) for each command during the first 1000 simulation steps
of the performance analysis test-case, by compiler, with the devirtualized iterators.}
\label{fig:tot-after}
\end{figure}

\begin{figure}
\includegraphics[width=\textwidth]{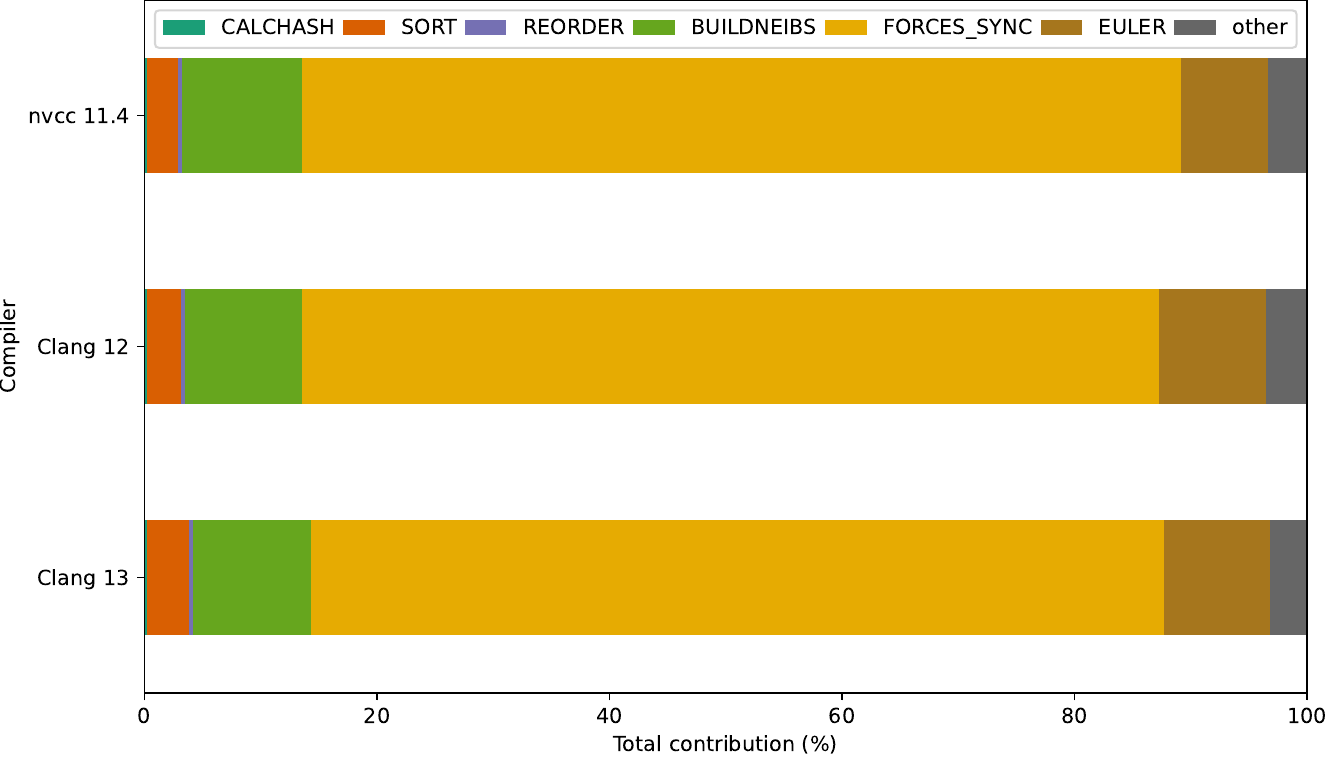}
\caption{Percent contribution by each command to the total runtime during the first 1000 simulation steps
of the performance analysis test-case, by compiler, with the devirtualized iterators.}
\label{fig:totpct-after}
\end{figure}

By using the \code{neibs\_iterator\_core} as the ``terminating'' class,
(i.e. the \code{NextIterator} when there are no more types to process),
we ensure that all iterators inherit a single copy of it as (grand)parent
through the chain, without having to resort to virtual inheritance.
By adding to \code{neibs\_iterator\_core} a \code{reset()} method that
invalidates \code{current\_type} and a \code{next()} method that always returns false,
the \code{nested\_neibs\_iterator} implements both single- and multi-type iterators
with the same code, and the only distinction is given by the nesting of the classes
(Figure~\ref{fig:chain-inheritance}).

As show in Figures~\ref{fig:avg-after}--\ref{fig:totpct-after}, 
the effect of the de-virtualization of the neighbors iterator on the performances is impressive:
the total runtime for \code{nvcc} and Clang~13 dropped from over 10s to less than 5s,
bringing these compilers in line with the performance of Clang~12,
that also benefits (although in much smaller amounts) from the optimization.
Moreover, without the bottleneck created by the problematic virtual inheritance,
the proprietary optimizations in \code{nvcc} allow the compiler to produce the fastest code
among the ones we tested.

\section{Too much of a good thing?}

The impressive performance gains achieved by the rework of the neighbors traversal,
that in some sense completes the split-neighbors implementation, have had some unintended consequences.

The first consequence has been an apparent decrease in the scaling efficiency of GPUSPH across multiple GPUs:
this is due to the fact that high scalability relies on the forces computation
taking enough time to cover the cross-device data transfer time
(see~\cite{rustico_multi-gpu_2014,rustico_advances_2014} for details):
with the forces computation being much faster,
the amount of inner particles that need to be processed to cover data transfers
has grown significantly, thus requiring much larger domains to achieve good scaling.
This of course is compensated by the fact that a single GPU now manages to perform as well as
4 or 5 GPUs used to with the older code, thereby reducing the need for multi-GPU or multi-node
to very large simulations, where the amount of inner particles becomes large enough
to preserve the scaling properties. Further details with examples are discussed in Section~\ref{sec:mgpu}

The second consequence of the forces kernels performance gain has more of a psychological than computational weight:
as shown in Figure~\ref{fig:avg-after},
forces computation is not the slowest step of the simulation anymore:
although forces computation is still the kernel where most time is spent overall
(Figure~\ref{fig:totpct-after}),
its runtime for a single invocation is now comparable to the runtime of the neighbors list construction.

In fact, for smaller simulations such as the one shown in this analysis,
a single run of the neighbors list construction can take more time
than a single forces computation, with the discrepancy growing smaller as the number of particle
increases, and forces computation still taking longer (although not much so) than neighbors list construction
in the case of very large simulations.

In itself this effect is not particularly worrying, since the forces kernel runs twice per step
(due to the predictor\slash corrector integration scheme), while the neighbors list construction
only runs once every 10 steps (by default). As such, even when the two have comparable runtime,
forces computation still weight 20 times more than the neighbors list construction, and optimization of the latter
would be largely unnecessary, since even a $50\%$ speed-up would only affect the total runtime
of the simulation by less than $3\%$.

Still there is a psychological effect: knowing that building the neighbors list takes longer
than a single forces computation is frustrating for a numerical code that \emph{should} spend
most of its time computing (numerical) derivatives.
Hence the need to optimize the neighbors list construction.

\subsection{Changing perspective: from the particle to the cell}

As mentioned in section~\ref{sec:sph-impl}, WCSPH with an explicit integrator lends itself
to a natural parallelization where each work-item is associated with a particle.
This is also true for the neighbors list construction: each particle looks for neighbors
in the cells surrounding its own (section~\ref{sec:nl-basis}), adding them to the neighbors list
according to their type (section~\ref{sec:split-neibs}).

This approach is trivial to implement, but leads to poor performance due to the scattered 
global memory accesses and consequent large latencies as the neighbors data is fetched,
with the work-items unable to proceed until the data becomes available and decisions
about the neighbors (Is it close enough? Which type does it belong to? Where does it go in the list?)
must be made.

The first optimization is to take advantage of the split-neighbors layout to streamline
the neighbors search. As particles are sorted in memory by cell,
and within the cell by particle type (and finally by particle ID),
we know that whenever looking at neighbors in a cell the central particle will first see
all fluid particles, then all boundary particles, etc.
Instead of a single loop going over all the particles in the cell,
we can thus split the code into multiple loops, one per particle type.

\begin{lstfloat}%
[caption={Traversing the neighbors in a cell by particle type.},
 label={l:neibsincelloftype}]
template<ParticleType nptype, /* omissis */ >
void neibsInCellOfType
	(params_t const& params, var_t const& var, /* omissis */)
{
	/* from the neighbor loaded in var, keep processing all neighbors
	 * until the next loaded one is not of the same type
	 */
	for ( ; var.is_neib_of_type<nptype>(params); var.next_neib())
	{
		/* process this neighbor */
	}
}

template<typename params_t, /* omissis */>
void neibsInCell(params_t const& params, /* omissis */)
{
	/* omissis */

	var_t vars(params, /* omissis */);

	// Load information about the first neighbor
	var.load_neib_info(params);

	// Process neighbors by type, leveraging the fact that
	// within cells they are sorted by type.
	// Each instance of neibsInCellOfType will process the neighbors
	// of the given type, and return when the next loaded neighbor
	// (if any) is of a different type
	if (var.neib_type == PT_FLUID)
		neibsInCellOfType<PT_FLUID, /* omissis */>
			(params, var, /* omissis */);

	if (var.neib_type == PT_BOUNDARY)
		neibsInCellOfType<PT_BOUNDARY, /* omissis */>
			(params, var, /* omissis */);

	if (boundarytype == SA_BOUNDARY && var.neib_type == PT_VERTEX)
		neibsInCellOfType<PT_VERTEX, /* omissis */>
			(params, var, /* omissis */);

};
\end{lstfloat}

Since the structure of the loop is largely the same, the loop can be implemented
as a function template, parametrized on the neighbor type,
taking care of all the neighbors of the given type, and returning when
the next neighbor is of a differen type (or there are no more neighbors).
The instances of the function can be called in sequence,
following the in-cell sort order
(Listing~\ref{l:neibsincelloftype}).
This change alone has been sufficient to give a performance gain in the order of $10\%$
for the neighbors list construction, due to the reduction in divergence
and to some control logic moving from runtime to compile time.

More significant improvements require a complete change in perspective.
To explain how, we need to observe what happens at the hardware level during the execution
of the standard implementation on GPU.

Consider two particles with consecutive indices. In many cases, they will belong to the same cell,
and process neighboring cells concurrently. In fact, with the per-particle perspective adopted so far,
they will process the same particles in the neighboring cells concurrently:
they will load at the same time the first particle in the cell at offset $(-1, -1, -1)$,
decide what to do with it, then move (at the same time) to the second particle in the same cell, etc,
until the neighboring cell is exhausted, and then move to the next cell.

This is not an ideal access pattern: even assuming that the memory controller supports
broadcasting (so the neighboring particle data is fetched for all work-items at the same time),
the effective bandwidth is limited, since each transaction only transfers the data of a single particle,
whereas the controller on GPUs is designed to transfer data from \emph{consecutive} memory locations
to \emph{consecutive-index} work-items (which is the reason why the neighbors list is stored
in interleaved format on GPU, as explained in section~\ref{sec:split-neibs}).

To maximize throughput, we would need work-items to transfer data corresponding to \emph{adjacent}
particles: for example, while the work-item for particle $i$ requests the data for the first particle
from the cell at offset $(-1, -1, -1)$, the work-item from particle $i+1$ should request the data
for the \emph{second} particle in the cell, and so on.
Of course, all work-items would then need the information from the neighboring cell particles,
which have been loaded by \emph{other} work-items: this data can be exchanged by storing it
on fast on-chip shared memory (accessible by all the work-items in the same work-group).

To maximize the usefulness of this approach, we can switch from a particle-based perspective
to a cell-based perspective, leveraging the fact that on NVIDIA GPUs work-items (a.k.a. \emph{CUDA threads})
do not execute independently, but proceed in lock-step grouped in 32-wide sub-groups (a.k.a. \emph{warps}).

The idea is thus to map each warp to a cell, and each work-item in the warp to a particle in the cell.
This \emph{guarantees} that all work-items in the same warp traverse the same neighboring cells, in lock-step.
Each work-item (from the same warp) then loads one particle from the neighboring cell into the shared memory,
and all work-items (in the warp) process the neighbors taking the data from the shared memory.

Compared to the naive thread-to-particle mapping, the warp-to-cell mapping is considerably more complex to implement.
One of the main difficulties is that it's not possible to do early bail-outs:
even work-items that are not associated to active particles must process the neighboring cells,
to ensure that all neighbors data is correctly loaded; this requires additional boolean variables
to be carried around, controlling whether the work-item must also process the neighbors, or only contributes to the
data retrieval.

Additionally, a cell may contain more particles than there are work-items in the warp.
Therefore, both the central particle selection and the neighbor data loads are run in a ``sliding window'' fashion,
until exhaustion of the cell particles.

\begin{figure}
\includegraphics[width=\textwidth]{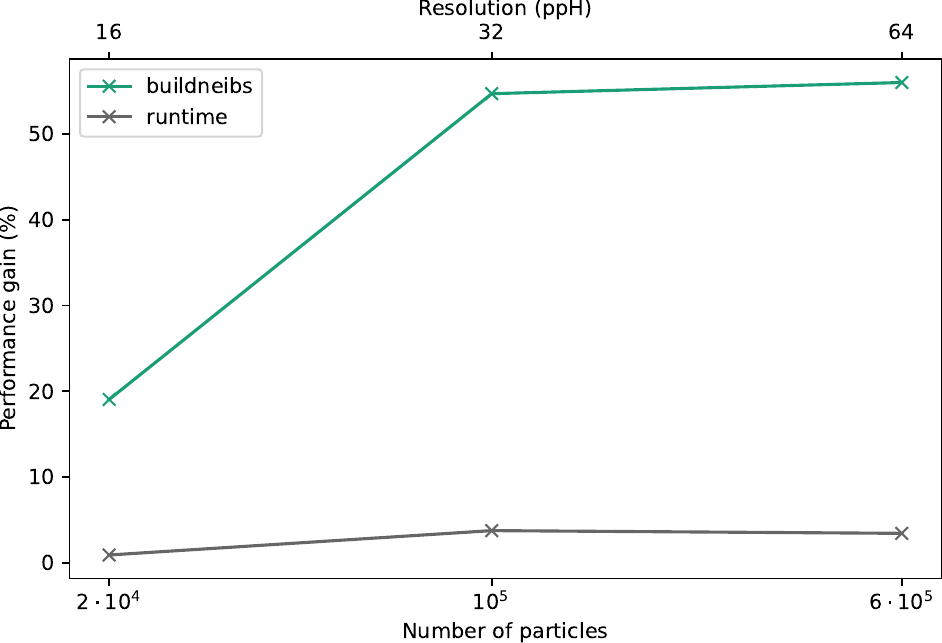}
\caption{Performance gain (see the text for the details) in the average runtime for the neighbors list construction
and in the total runtime, using the per-cell approach over the per-particle approach, with the \code{nvcc} compiler.}
\label{fig:avg-neibs}
\end{figure}

That being said, in the most common configuration of SPH smoothing kernels with radius $k = 2$
and smoothing factors of $h = 1.3$, cells in three-dimensional problems have less than 32 particles.
This actually leads to a sub-optimal usage of the hardware, since every running warp will have some work-items
masked out (typically, only the first 20 to 24 work-items in each warp will be active), leading to
a hardware efficiency of $75\%$ \emph{or less}, compared to the particle-based approach
where all work-items will be active and running.

Despite this inefficiency, however, the cell mapping comes out to be around $50\%$ \emph{faster} than the
particle-based approach, due to the significant improvements in memory access patterns and the consequent reduction in latency.
This is consistent with Volkov's findings about the relative importance of latency and occupancy
in gauging parallel algorithms and their performance on GPU~\cite{Volkov_talk,Volkov_phd}.

This performance gain (computed as the relative change in runtime, i.e. $T_{\text{old}}/T_{\text{new}} - 1$,
where $T_{\text{old}}, T_{\text{new}} - 1$ are the runtimes of the old and new code)
in the neighbors list construction runtime only has a modest effect on the total runtime
(Figure~\ref{fig:avg-neibs}), diminishing as the number of particles grows larger.

As mentioned, this is expected, due to the lower frequency of execution of the neighbors list construction
in typical simulations, but the benefits can become important if the list needs to be updated more frequently.

\section{An industrial test case}\label{sec:benchmarks}

The performance gains presented so far are crucial in industrial applications of SPH,
for which hundreds of millions of particles are typically employed, due to large domain sizes and\slash or
fine resolutions.

We illustrate this in a real-world test case, showing performance results including single-node
and multi-node scaling performance before and after the recent optimizations,
and validation of the simulation results against experimental data.

\subsection{Test case setup}

\begin{figure}
\includegraphics[width=\textwidth]{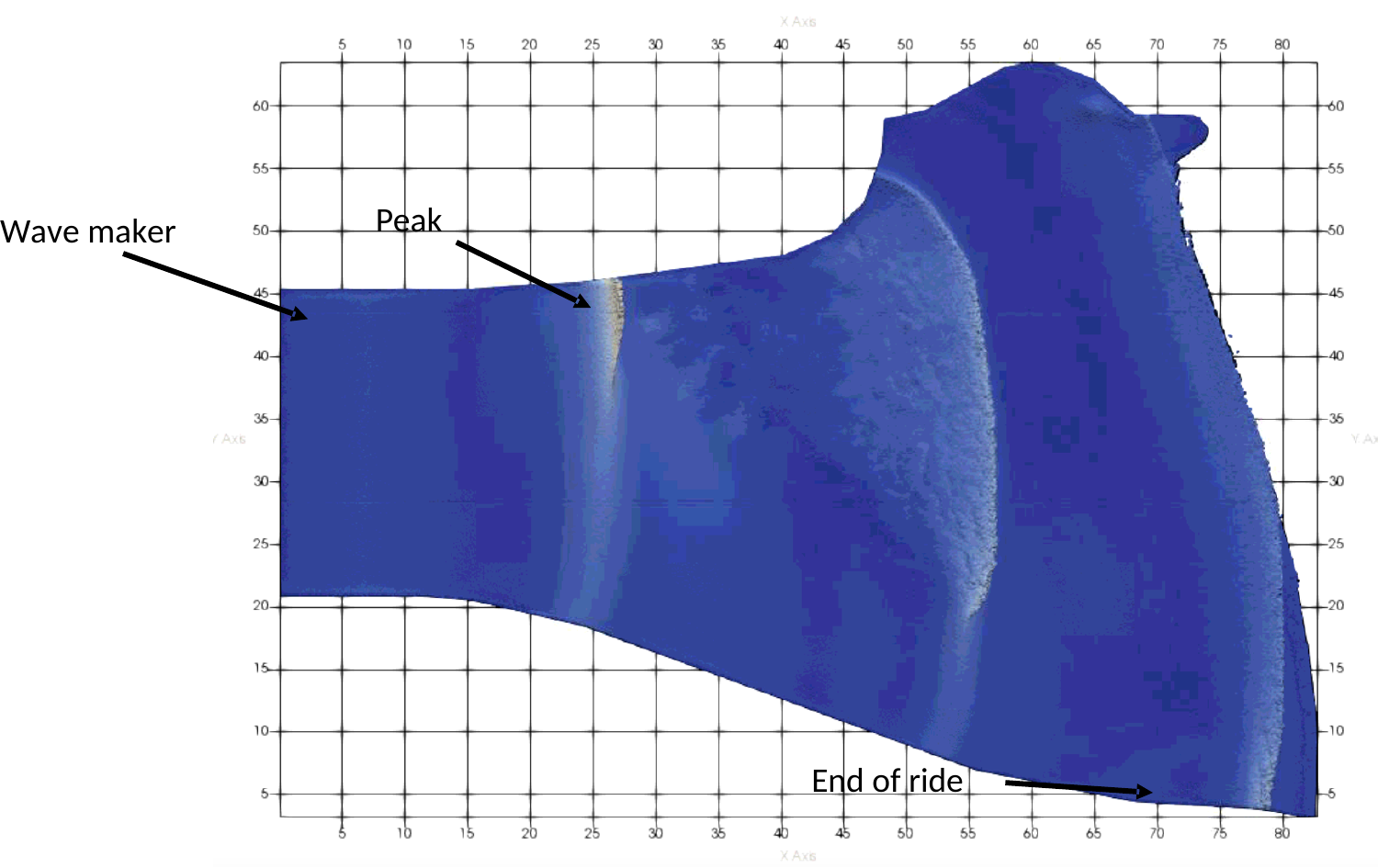}
\caption{Overhead view of the irregularly-shaped wave basin with variable bottom bathymetry used in the test case,
including the location of the three wave gages used for the comparison between the experimental and numerical data.}
\label{fig:domain}
\end{figure}

The test case refers to a large-scale wave maker and basin with a bounding box that is
approximately $65\text{m}\times85\text{m}$ (Figure~\ref{fig:domain}).
The wave is generated by a caisson system to produce a nominal maximum wave height $H = 0.9$m
with period $T=7.5$s.
The water depth at the caisson is $d = 3.4$m and decreases to $0$m at the end of the basin
with a non-uniform reef bathymetry in-between.

For the SPH simulations, the Lennard--Jones boundary model was used,
with a geometric description of the basin bathymetry through a Digital Elevation Model
exerting a point-wise normal Lennard--Jones force~\cite{bilotta_2016}.
The water is assumed inviscid, and no turbulent or artificial viscosity terms are included in the momentum equation.
The Molteni \& Colagross density diffusion model~\cite{molteni_colagrossi_2009} is used.
The chosen SPH kernel is the Wendland quintic with radius~2, and smoothing factor~1.3.

The hardware used for the run consists of 3~nodes with 3 RTX 3090 GPUs each.
Each GPU is equipped with $24$GB of GDDR6X VRAM running at $1.2$GHz wit a peak theoretical bandwidth of $936$GB/s,
and $82$ compute units with $128$ shader processors each for a total of $10,496$ ``CUDA cores'' per GPU,
with a peak theoretical throughput of around $30$TFLOPS for single-precision FMA.
Due to the hardware setup, peering between GPUs within a node was unavailable,
lowering the multi-GPU scaling performance, as discussed below.
The nodes are equipped with 1Gbps network cards and are interconnected through a 24-port switch.

\subsection{Validation}

\begin{figure}
\centerline{\includegraphics[width=\textwidth]{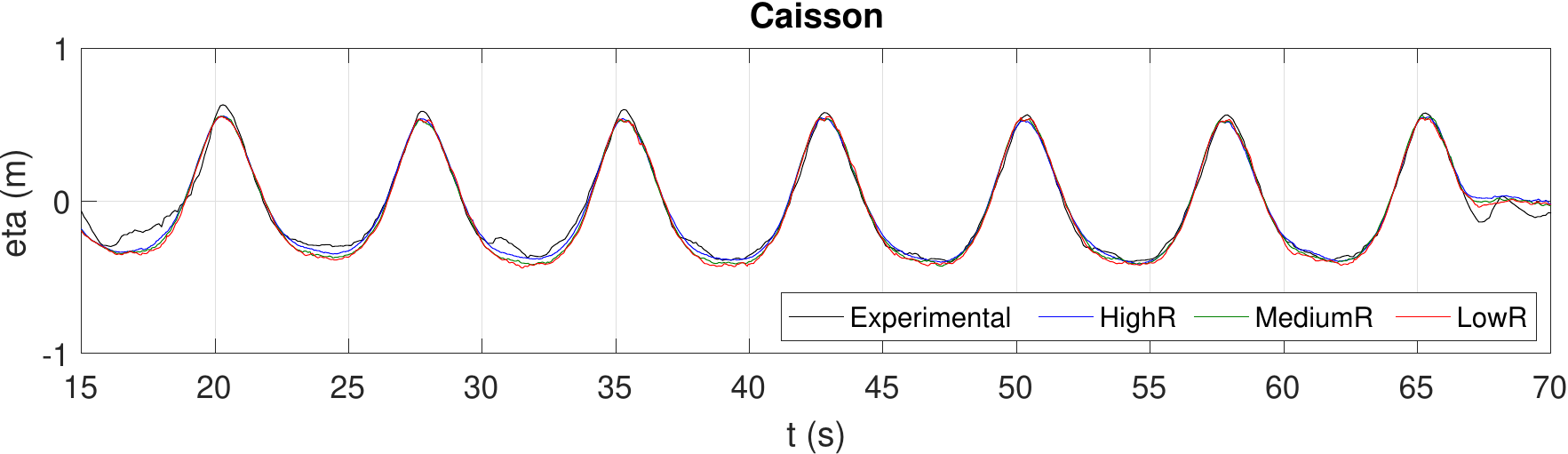}}%
\centerline{\includegraphics[width=\textwidth]{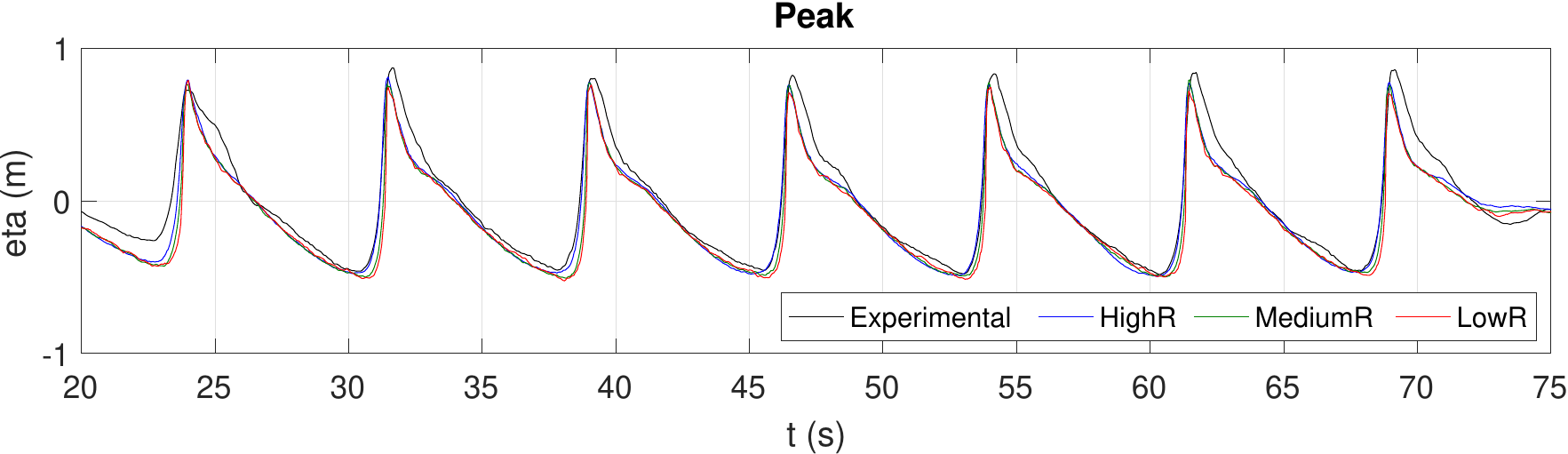}}%
\centerline{\includegraphics[width=\textwidth]{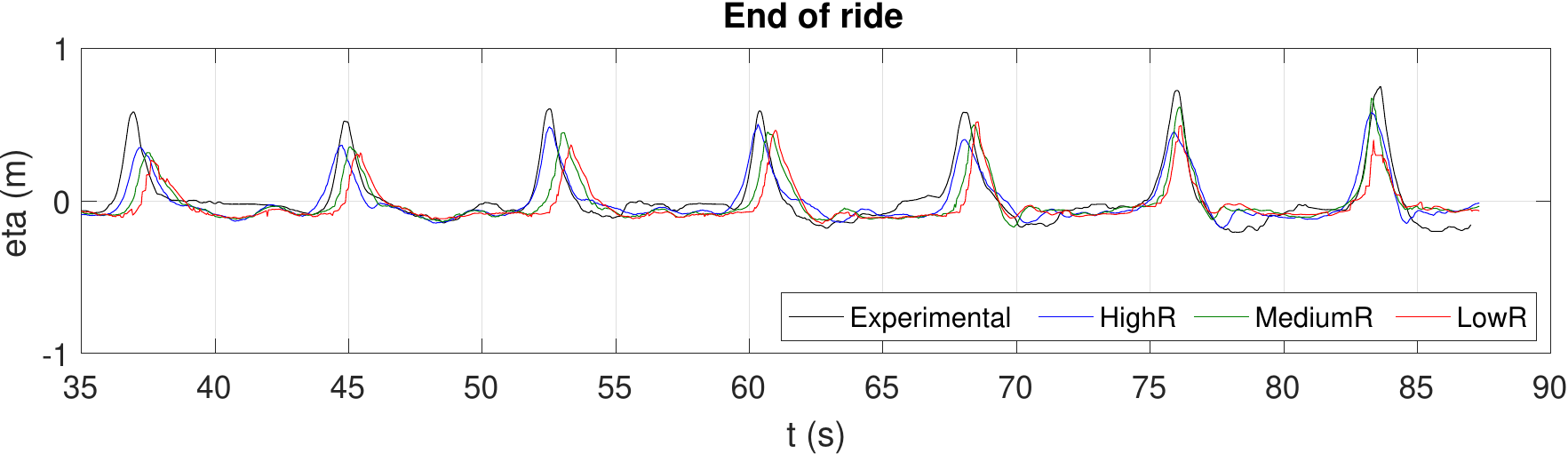}}%
\caption{Water surface elevation over time from the experimental data compared with the low, medium and high resolution SPH simulations
at the caisson (top figure), peak (middle figure) and end-of-ride (bottom figure).}
\label{fig:expplots}
\end{figure}

Comparison between the physical experiments and SPH simulations were done comparing the water surface elevation and time-of-arrival
at three distinct gage positions: near the caisson, where the peak wave height is expected, and at the end of the ride
(Figure~\ref{fig:expplots}).
Simulations were run at thee different resolutions, with inter-particle spacing respectively $\Delta p = 0.01, 0.008, 0.005$m
resulting respectively in (fluid $+$ boundary $=$ total) $4 + 2.4 = 6.4, 7.7 + 3.8 = 11.5$ and $32 + 10 = 42$ million particles.

Even though the chosen options result in a quite simple model, the results show good accuracy.
Thanks to the choice of density diffusion model, the simulation is stable over the simulated period
despite the absence of viscous dissipation in the momentum equation.
As expected, the match of the wave timing and crest height improve at higher resolutions.
The largest discrepancy is seen at the end-of-ride gage, where excessive dissipation is observed even at higher resolutions.
This effect may be reduced with the adoption of better conservation models such as the CCSPH model presented in~\cite{zago_ccsph},
at the cost of higher computational requirements.

\subsection{Performance results}

\begin{figure}
\includegraphics[width=.49\textwidth]{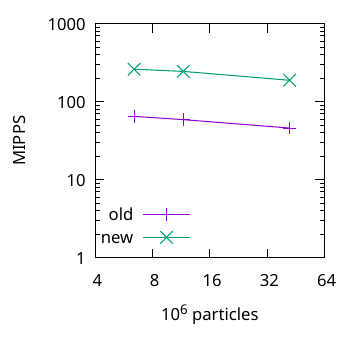}\hfill
\includegraphics[width=.49\textwidth]{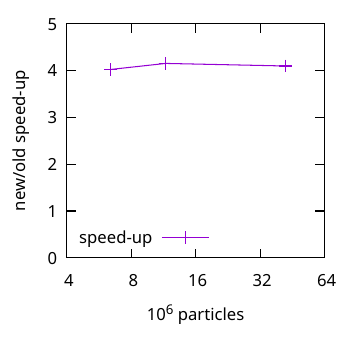}
\caption{Single-GPU performance with the old (unoptimized) and new (optimized) neighbors list construction and traversal code.
Left: performance (in MIPPS). Right: performance ratio.}
\label{fig:new-single}
\end{figure}

\paragraph{Single-GPU}
In our performance analysis we start by comparing the single-GPU performance across different resolutions
before and after introducing the neighbors list construction and traversal optimizations.
The performance is used in millions of iterations times particles per second (MIPPS),
a metric we first introduced in~\cite{rustico_multi-gpu_2014}.
As show in Figure~\ref{fig:new-single}, the performance ratio for this test case at the tested resolutions
is essentially constant, with the optimizations bringing a consistent $4\times$ speed-up to the code.
The decrease in MIPPS at higher resolutions is due to the higher percentage of fluid particles in the total particle count,
and the higher computational requirements for fluid particles, for which the physical equations of motions
have to be solved, over boundary particles, that only contribute to the forces applied to the fluid particles in their influence sphere.

\paragraph{Multi-GPU}\label{sec:mgpu}

We can evaluate the strong scaling capability of GPUSPH by comparing the performance of our code at given resolutions
on a growing number of device.
This is usually achieved by looking at the ratio $T_1/(n T_n)$ where $T_i$ represents the runtime with $i$ devices.
Using our preferred unit of measure is the number of iterations times particles per second,
the scaling efficiency can be likewise measured as $P_n/(n P_1)$ where $P_i$ is the number of MIPPS achieved with $i$ devices.

Due to the embarrassingly parallel nature of WCSPH with explicit integration ---as used in our tests---
the only major factor impacting scalability comes from the latency introduced between solver steps
when transferring data about neighbors to\slash from other devices.
As explained in~\cite{rustico_multi-gpu_2014,rustico_advances_2014}, each device in GPUSPH holds a copy
of the data belonging to neighboring particles (``halo'' particles) that reside on other device, and this is updated
after each computational kernel, with the exception of the integration, which is a simple increment
without any neighbors list traversal, and is therefore done on the halo particles
with the forces data transferred from the neighboring devices after the forces computation kernel.

Historically, the forces computation kernel has been the most computationally expensive kernel,
and this has been used to minimize the impact of data transfer on scaling performance,
by first computing the forces on the particles to be transferred to other devices,
and then running the computation on the rest of the domain (inner particles)
while the computed forces are being copied.
This allowed GPUSPH to achieve nearly ideal scaling~\cite{rustico_multi-gpu_2014,bilotta_bicgstab}
under the condition that the time needed to transfer the halo particles data
is less than the time needed to compute the forces on the inner particles.

This typically requires fast device-to-device transfers (e.g. through peering for devices on the same node,
or solutions such as GPU Direct in multi-node configurations), and a sufficient number of inner particles.
With the speed-up of the forces kernel computations coming from the optimized neighbors traversal code,
we can expect the scaling performance of the new code to degrade compared to the scaling performance
of the old code, especially when the number of particles (and especially inner particles) is small.
The effect will be particularly evident in our tests due to the hardware configuration that prevents
device peering, and even more so in the multi-node cases due to the higher latency and lower bandwidth
of network data transfers compared to transfers within the same node.

\begin{figure}
\includegraphics[width=\textwidth]{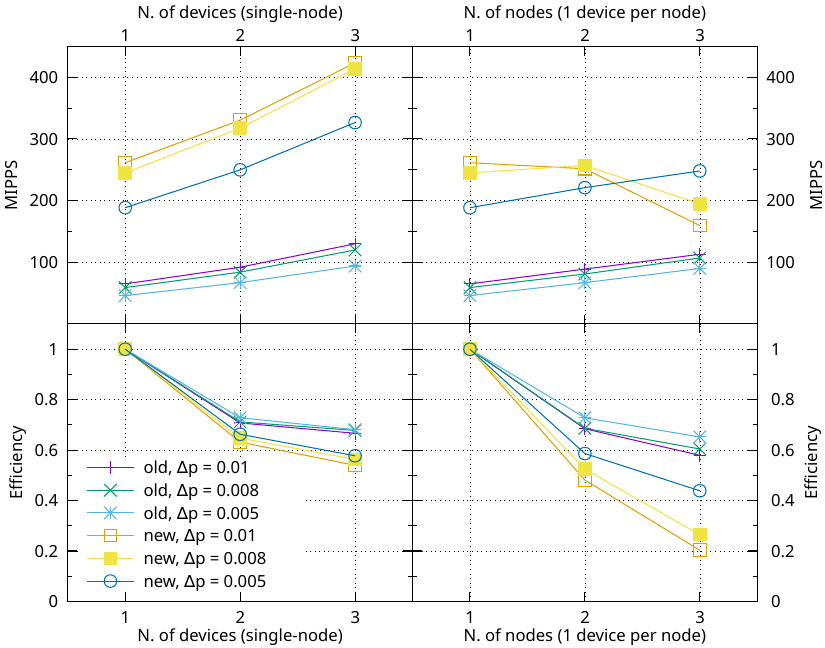}
\caption{Performance (top) and strong scaling (bottom) at different resolutions.
The left plots refer to the multi-GPU single-node case.
The right plots refer to the multi-node case, with 1~GPU per node.}
\label{fig:scaling13}
\end{figure}

The results for $1$ to $3$ devices are shown in Figure~\ref{fig:scaling13},
separating the single-node multi-GPU results (with all devices attached to the same host machine)
from the multi-node results, limited in this case to 1~GPU per node.

In the single-node case, where data transfers are more efficient,
all tests show performance growing with the number of GPUs.
The old code has lower performance, but better scaling,
with an efficiency of around $70\%$, whereas the new code drops
to an efficiency as low as $53\%$ in the low resolution case
($57\%$ at the highest resolution).

This is due to the much faster forces computation kernel in the new version of the code
failing to fully cover the latency introduced by the data transfer.
The effect is even more evident in the multi-node, 1~GPU per node case,
where the overall performance of the code actually \emph{drops}
as the number of nodes grows, with the only exception being the highest resolution case,
where the workload for the forces computation kernel is sufficient to
cover at least part of the network data transfers.
Both versions of the code suffer from the more expensive data transfers,
with the old code efficiency dropping to between $58\%$ (low resolution)
and $65\%$ (high resolution), and the new code
only managing $44\%$ efficiency at high resolution, with the low resolution dropping to~$20\%$.

\begin{figure}
\includegraphics[width=.49\textwidth]{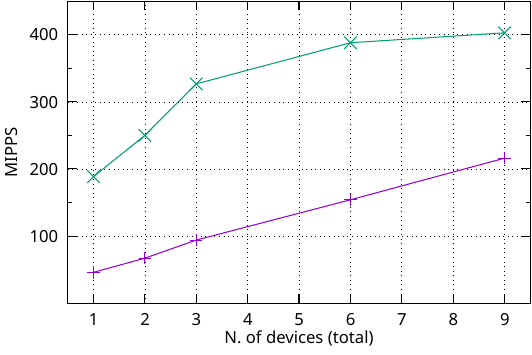}\hfill
\includegraphics[width=.49\textwidth]{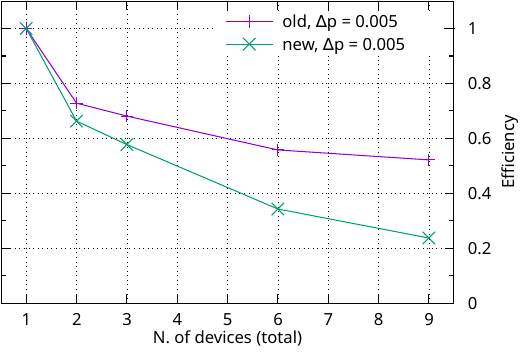}
\caption{Performance (left) and strong scaling (right) for the high resolution case.
A single node is used for up to 3 devices, with 6 and 9 devices using~2 and~3 nodes respectively.}
\label{fig:scaling69}
\end{figure}

Due to the insufficient computational load in the lower resolution cases,
we only test the configuration with up to 9 devices (3 nodes with 3 devices each)
in the high resolution case.
The results are shown in Figure~\ref{fig:scaling69}.
Increasing the number of devices in this case always results in higher performance
(and thus faster simulations), but there are diminishing returns as the number of particles
assigned to each device decreases.

Once again the effect is more evident with the new, faster code, that needs a larger number of particles
for the forces kernel computation to take enough time to cover the data transfers,
especially in the multi-node, 3~GPUs per node configurations. Given the scaling of the forces kernel,
we would expect no less than $6\times$ the number of inner particles per GPU to be needed to achieve scaling efficiency comparable to that of the old code.

\section{Conclusions}

The more recent revisions of the C++ language provide opportunities
that can help in the design and implementation of complex code bases
for scientific computing that need to support several features without
loss of performance and keeping code complexity low.

Key programming patterns and C++ features such as conditional inheritance for class templates
and SFINAE for function template specialization allow significant levels of code abstraction
and genericity, providing at the same time several opportunities for the compiler to optimize 
away unnecessary variables and computations.

Helping the compiler produce better code is an effective way to speed up your program,
but the final result still largely depends on the quality of the compiler,
and the ability of its optimizer to leverage the opportunities offered by the developer.

On GPUs, where some programming patterns can result in excessive use of the main memory
instead of hardware registers to hold temporary data, catching or missing optimization opportunities
can change the performance of individual kernels by nearly an order of magnitude.
As a result, the choice of the compiler can be at least as important as the coding strategies
in delivering the best performance on any given hardware,
although significant differences between compilers are a likely indication of ``code smells'',
i.e. sections of the code that are still using sub-optimal, and thus harder to optimize, patterns.

When it comes to implementations of the SPH method, and meshless methods in general,
one of the crucial aspects in the performance of the code is the neighbors list:
its memory layout, the way it is traversed and the way it is constructed.
Careful design of the data structure and traversal methods can lead to performance gains
(or losses)
measurable in a factor of $3\pm1$, especially on hardware such as GPUs that are exceptionally sensitive
to register spilling and inefficient global memory usage patterns.

A paradox with the performance increase is a potential drop in the strong scaling efficiency in multi-GPU,
and especially multi-node, simulations, as the shorter execution time can fail to fully cover
the time necessary to transfer data between devices.
This is more than compensated by the shorter computational times in general,
and may imply that a much higher per-device load may be necessary to achieve comparable efficiency.
Hardware setups with higher connection speed and lower latency can also help improve scaling.

The neighbors list construction itself, or in its absence the neighbors search implementation,
also has a measurable impact, weighted by the frequency at which it is run.
As an intrinsically memory-bound procedure, neighbors search can benefit from a change in perspective
that leads to improved coalescence of memory transactions and better caching on GPU.
In our experience, switching from the naive per-particle vision to an aggregate per-cell approach
can lead to performance gains of as much as $50\%$ in the neighbors search despite a lower hardware occupancy.
It's possible that similar optimizations may also be possible for the main computational kernels
implementing the actual numerical method,
indicating that despite the appeal of SPH as an embarrassingly parallel method,
optimal performance lies behind non-trivial implementation strategies.

These kinds of transformations of the numerical method into efficient code are still far from the reach of the compiler,
keeping human developers still as the most effective optimizers.
Despite the progress in compiler technology,
there is only so much the compilers can do on their own, and it thus remains crucial for the developer
to provide as much information as possible to the compiler itself,
reorganizing both data structures and their access functions appropriately.
The programming strategies presented in this paper,
by moving as much of the control logic as possible to the compile time
and by streamlining data dependencies,
remain therefore essential in achieving the best performance the compiler can offer,
at the cost of
a less obvious mapping between the theory of the numerical method and the code implementing it,
and longer development times.
The latter, however, may be more than compensated by the performance gains of large simulations
(frequently taking months to complete themselves) when speed-ups such as the ones obtained in our case
are possible: runtimes dropping from months to weeks are definitely worth the investment in a month of developer time.
Additionally, the abstractions and code strategies presented here are key elements
to the implementation of heterogeneous hardware support, covering CPUs and GPUs from different vendors,
as shown in the upcoming manuscript~\cite{bilotta_cpu_gpusph}.





\bibliographystyle{elsarticle-harv}
\bibliography{gpusph.bib,nongpusph.bib}







\end{document}